\documentclass[lettersize,journal]{IEEEtran}
\usepackage{tikz}
\usepackage{amsmath}
\usepackage{filecontents}
\usepackage{diagbox}
\usepackage{subfigure}
\usepackage{algorithm,algorithmic}

\usepackage{amsmath}
\usepackage{amssymb}
\usepackage{multicol}
\usepackage{threeparttable}
\usepackage{rotating}
\usepackage{makecell}
\usepackage{multirow}
\usepackage{url}
\usepackage{hhline}
\usepackage{booktabs}
\usepackage{pifont}
\usepackage{mathrsfs}
\usepackage[many]{tcolorbox}
\usepackage{tikz}
\usepackage[caption=false,font=normalsize,labelfont=sf,textfont=sf]{subfig}
\usetikzlibrary{fit}

\begin{document}
\date{}
\title{PKDGA: A Partial Knowledge-based Domain Generation Algorithm for Botnets}
\vspace{-55pt}
\author{\IEEEauthorblockN{Lihai Nie\IEEEauthorrefmark{2},
Xiaoyang Shan\IEEEauthorrefmark{2},
		Laiping Zhao\IEEEauthorrefmark{2},
		Keqiu Li\IEEEauthorrefmark{2}}	\\ \IEEEauthorblockA{\IEEEauthorrefmark{2} Tianjin Key Lab. of Advanced Networking (TANKLab),\\ College of Intelligence \& Computing (CIC), Tianjin University, Tianjin, China}
\vspace{-27pt}
\thanks{L. Nie, X. Shan, L. Zhao and K. Li are with the College of Intelligence \& Computing, Tianjin University, Tianjin 300000, China.\protect\\
	E-mail: \{nlh3392, sxy2fj, laiping, keqiu\}@tju.edu.cn.}}
	
\markboth{Manuscript submitted to IEEE Transactions on Information Forensics and Security}%
{Shell \MakeLowercase{\textit{et al.}}: Bare Demo of IEEEtran.cls for IEEE Communications Society Journals}
\maketitle


\begin{abstract}
Domain generation algorithms (DGAs) can be categorized into three types: \emph{zero-knowledge}, \emph{partial-knowledge}, and \emph{full-knowledge}.
While prior research merely focused on \emph{zero-knowledge} and \emph{full-knowledge} types, we characterize their anti-detection ability and practicality and find that \emph{zero-knowledge} DGAs present low anti-detection ability against \textit{detectors}, and \emph{full-knowledge} DGAs suffer from low practicality due to the strong assumption that they are fully \textit{detector}-aware. 
Given these observations, we propose \emph{PKDGA}, a partial knowledge-based domain generation algorithm with high anti-detection ability and high practicality.
\emph{PKDGA} employs the reinforcement learning architecture, which makes it evolve automatically based only on the easily-observable feedback from \textit{detectors}. 
We evaluate \emph{PKDGA} using a comprehensive set of real-world datasets, and the results demonstrate that it reduces the detection performance of existing \textit{detectors} from $\textbf{91.7\%}$ to $\textbf{52.5\%}$. We further apply \emph{PKDGA} to the \textit{Mirai} malware, and the evaluations show that the proposed method is quite lightweight and time-efficient.  
                    


\end{abstract}
\vspace{-4pt}

\vspace{-8pt}
\section{Introduction}\label{sec:intro}
\vspace{-2pt}
\IEEEPARstart{D}OMAIN generation algorithms (DGAs) 
have been extensively adopted by modern botnets through generating a large number of domain names (i.e., algorithmically generated domains, AGDs~\cite{AGD_refer})) and then using a subset of those names for actual command and control (C\&C) communication~\cite{Khaos,Kraken,DGArchive,Gozi,Pykspa,Suppobox}. This ability makes it difficult to track communications with C\&C servers operated by the attacker, enabling the botnets to resist blacklisting and related countermeasures such as takedown efforts \cite{take-down-1, bot_take,FANCI, TDSC-takedown}. The DGA-enabled botnets can be used in a variety of cyber attacks, e.g., ransomware~\cite{Ransomware}, spam campaigns~\cite{spam-campaigns}, and distributed denial-of-service (DDoS) attacks~\cite{DDoS}.



\begin{figure}
	\vspace{-5pt}
	\centering
	\includegraphics[width=0.5\textwidth]{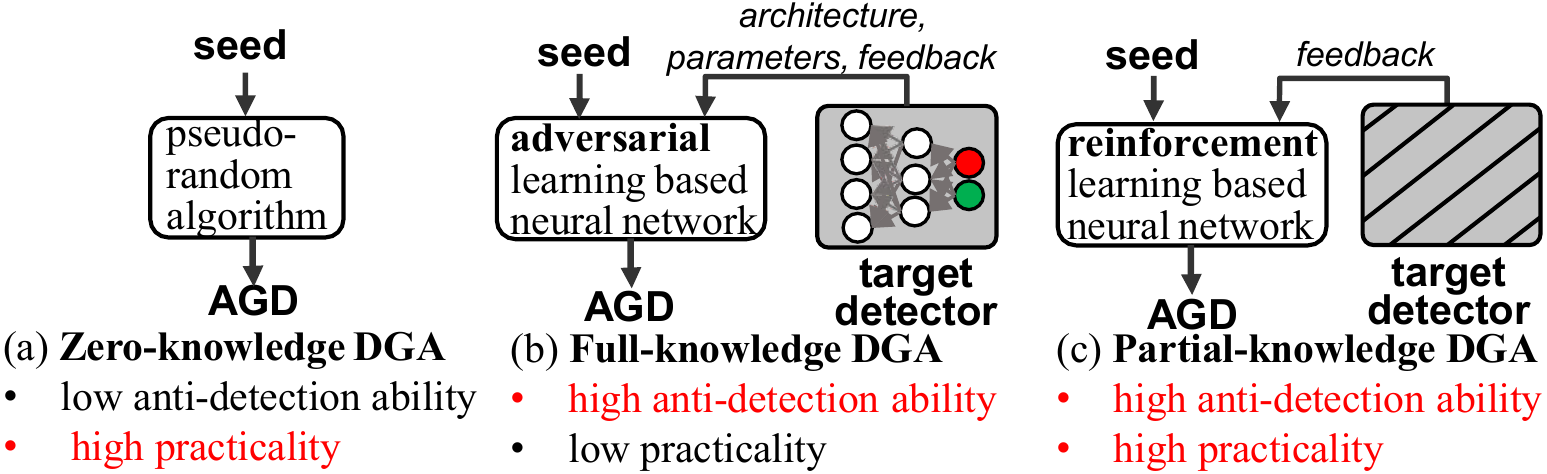}
	\vspace{-20pt}
	\caption{DGA classification in terms of knowledge requirements}
	\label{fig:DGA_cate}
	\vspace{-20pt}
\end{figure}

Prior DGAs typically sample elements from predefined dictionaries (e.g., wordlists, alphabets and ASCII tables) through seed-triggered pseudorandom algorithms and then concatenate the elements as a synthetic domain (i.e., the \emph{zero-knowledge DGA} in Fig.~\ref{fig:DGA_cate}(a) \cite{Kraken,Gozi,Pykspa,Suppobox}).
Seeds are the secrets shared by adversaries and victims. Once a DGA and its respective seed are known, the botnet can be easily hijacked (i.e., sinkholing) and adversaries can then redeploy new botnets with updated seeds.
As a countermeasure, security vendors have developed machine learning (ML)-based \textit{detectors} to identify AGDs generated by zero-knowledge DGAs through analyzing their distinctive features like graphical~\cite{Graph_detection}, linguistical~\cite{FANCI} and statistical characteristics~\cite{Distance-based}. Evaluations show that they can achieve $\ge$$90\%$ accuracy in identifying the wordlist, alphabet and ASCII table-based DGAs \cite{Gozi,Suppobox,Kraken}.


Since ML-based \textit{detectors} are easily compromised by adversarial samples (i.e., maliciously crafted inputs cause the ML models to predict incorrectly~\cite{adversarial_example_1,adversarial_example_2, AWA}), adversaries further design \emph{full-knowledge DGAs}, which train a neural network-based adversarial AGD generator by assuming the full knowledge of \textit{detectors}~\cite{Khaos,MaskDGA}. As illustrated in Fig.~\ref{fig:DGA_cate}(b),
while the \textit{detector} architecture, model parameters, feedback as well as potential vulnerability are exposed, 
DGA models get updated via adversarial learning against the \textit{detector} and generate adversarial AGDs that are hardly identified by \textit{detectors}.



However, although \emph{full-knowledge DGAs} improve the anti-detection ability, the assumption of \emph{full \textit{detector} knowledge-aware} is strong, making it hard to apply them in the real world. In particular, adversarial AGDs cannot be obtained as long as the \textit{detector} architecture and its model parameters are not given in advance. To achieve both high anti-detection ability and practicality, we explore novel DGAs by reducing the acute dependence on \textit{detector} information while keeping the anti-detection ability (aka "\emph{partial-knowledge DGA}" in Fig. \ref{fig:DGA_cate}(c)).



\vspace{-1pt}
A \emph{partial-knowledge DGA} introduces several challenges that must be addressed. 
(1) \emph{What knowledge can be obtained easily that is also useful for DGA design}? While the \emph{model architecture} and its \emph{parameters} are often unavailable to adversaries, the \emph{detection feedback} can instead be easily observed by trying to register the testing domains. Therefore, it is possible to make use of the \emph{detection feedback} solely to improve the anti-detection ability of DGAs. 
(2) \emph{How to use such information to design a DGA with high anti-detection ability}? In theory, DGAs should be able to evolve based on the \emph{detection feedback} to improve the anti-detection ability. The reinforcement learning (RL) architecture is well-suited for such a design because it uses the experience gained through interacting with the \textit{detector} and evaluative feedback to improve the anti-detection ability to make behavioral decisions. 
(3) \emph{How can DGAs be integrated into attack tool kits while maintaining low overhead}? We evaluate the overheads of the existing \emph{full-knowledge DGAs} by implementing them in real malware (\textit{Mirai}~\cite{Mirai}) and find that they consume $30\times$ more memory than that without adopting DGA. To reduce memory overhead, we need to optimize the implementation by eliminating unnecessary libraries and runtime components.

\newpage

To overcome the aforementioned challenges, we build \emph{PKDGA}, a domain generation algorithm that achieves both high anti-detection ability and high practicability.
\emph{PKDGA} employs the RL architecture, which can automatically evolve based on the feedback from \textit{detectors}. Therefore, it enables adversaries to compromise arbitrary malicious domain detection systems and help botnets to avoid the domain-related detection and takedown efforts using only observable feedback. 
It is also lightweight in terms of memory and CPU overhead, facilitating the deployment of real malware with small footprints.

The contributions of this paper are summarized as follows.
\begin{itemize}
	\item We observe that the AGDs generated by existing DGAs can either be easily identified by prior \textit{detectors} or strongly rely on \textit{detector} knowledge during their generation process, which limits their application in botnets. 
	
	\item We demonstrate that high practicability and high anti-detection ability can be simultaneously achieved through the RL architecture. We design \emph{PKDGA}, that transforms the domain synthesizing task into a token sequence generation problem and optimize it using RL paradigm. 
	
	\item We comprehensively evaluate \textit{PKDGA} using a broad set of benign and malicious domains. The results show that \textit{PKDGA} can reduce the Area Under the Curve (AUC) of state-of-the-art \textit{detectors} from $91.7\%$ to $52.5\%$.

	\item We implement \emph{PKDGA} with open interfaces and apply it to the real malware \textit{Mirai}. The performance evaluations show that \emph{PKDGA} is quite memory efficient and that its inference time is also very short. 
\end{itemize}

The rest of the paper is structured as follows. We study the existing AGD \textit{detectors} and DGAs (\S \ref{sec:pre}), and use the findings to guide the design (\S \ref{sec:des}) and implementation (\S \ref{sec:imp}) of \emph{PKDGA}. We then evaluate \textit{PKDGA} (\S \ref{sec:eva}), discuss (\S \ref{sec:dis}), summarize related work (\S \ref{sec:rel}), and conclude (\S \ref{sec:con}).

 \vspace{-8pt}
\section{Preliminaries}\label{sec:pre}
 \vspace{-10pt}
\subsection{Backgrounds}

\begin{figure}[!htb]
	\vspace{-10pt}
	\centering
	\includegraphics[width=0.495\textwidth]{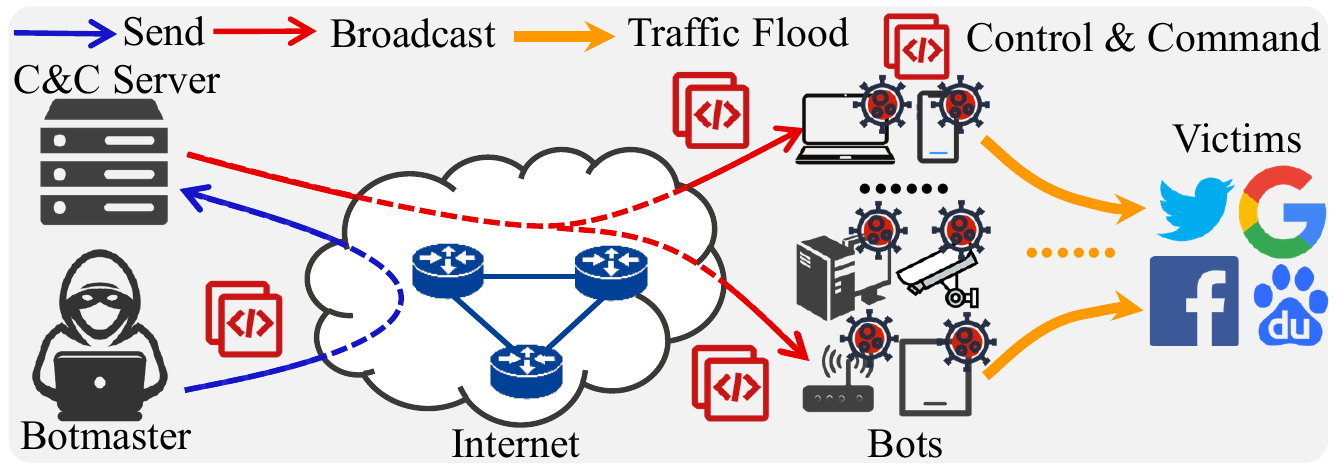}
	\vspace{-20pt}
	\caption{Botnet Illustration}
	\vspace{-8pt}
	\label{fig:overview1}
\end{figure}

\subsubsection{Botnet}
As illustrated in Fig.~\ref{fig:overview1}, a botnet is a network that consists of a \textit{botmaster}, a \textit{control and command} \textit{(C\&C) server}, and several \textit{bots}. Bots refer to the malware-infected networked devices that run bot executables and process the tasks supplied by the botmaster. The botmaster is the owner of the botnet, and they are capable of remotely controlling and commanding all of the bots to cooperatively attack the target victims. Since revealing their identity can lead to potential prosecution, the botmaster attempts to avoid detection by controlling all of the bots via a C\&C server. Here, the C\&C server is a relay node that receives the controls and commands from the botmaster and subsequently broadcasts the received messages to the bots.

Once a networked device is infected by malware, it becomes a potential bot and attempts to connect with the C\&C servers to join the botnet. 
It runs bot executable to derive the C\&C server's domain and then resolves that domain to obtain the IP address. However, a botnet can be easily taken down by blacklisting the domain hard-coded in the bot executable, as bots will fail to resolve C\&C server's domain successfully.

\newcommand{\whiteding}[1]{\raisebox{0pt}{\textcircled{\raisebox{-0.8pt} {\small\textsf{#1}}}}}
\newcommand{\whitedings}[1]{\raisebox{0pt}{\textcircled{\raisebox{-0.5pt} {\footnotesize\textsf{#1}}}}}
\subsubsection{Domain fluxing}
\begin{figure}
	\vspace{-12pt}
	\centering
	\includegraphics[width=0.49\textwidth]{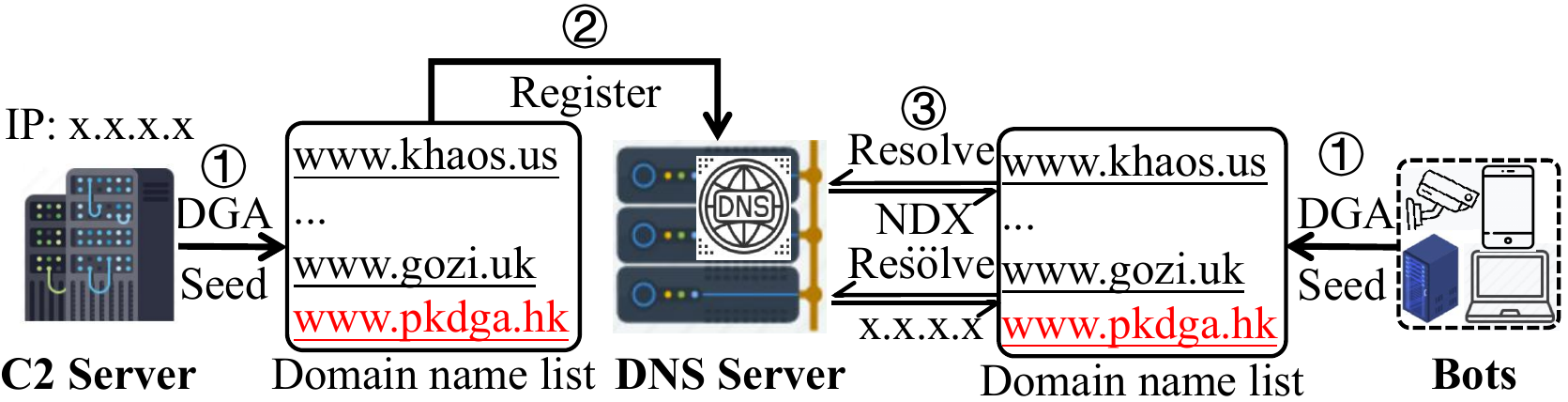}
	\vspace{-20pt}
	\caption{Domain fluxing}
	\label{fig:domain_fluxing}
	\vspace{-18pt}
\end{figure}

To resist take-down efforts that block static domains, domain fluxing technique is proposed. Specifically, bots communicate with the C\&C server using the DGA-generated domains, which are periodically updated. As shown in Fig.~\ref{fig:domain_fluxing}, the connections between the bots and the C\&C server are established through the following steps: \whiteding{1} Both the bots and the C\&C servers produce identical domain name candidates by running the same DGA using the same seed. 
\whiteding{2} The C\&C server registers a small fraction or even only one of the generated domain names with its IP address. \whiteding{3} At this time, the bots are still unaware of which domains have been registered, and they must resolve iteratively all of the candidate domains until the C\&C server's IP is successfully returned.
The bots connect the C\&C server using the resolved IP and receive control and command messages from the server.

\subsubsection{Domain Generation Algorithms (DGAs)} DGAs are the key techniques used in domain fluxing, and they are devoted to dynamically providing identical domain name(s) for both the bots and the C\&C server. There are two components in a DGA: a seed and a generation algorithm. A seed is a predefined dynamic string (e.g., current date or trending topics on \textit{Twitter}) that is consistently identical for both the bots and the C\&C server at any time. Given a seed, the generation algorithm can then produce domains. Moreover, the generation algorithm is required to be pseudorandom, that is, it yields identical domain names once the seed is deterministic. Thus, the bots and the C\&C server always have access to the same candidate domain name(s) when using DGA techniques.

\subsubsection{AGD detection}
Algorithm-generated domains (\textit{AGDs}) refer to the domain names yielded by DGAs and maliciously registered with the C\&C server's IP. Accurately identifying them facilitates the take-over or take-down efforts against botnets.
The AGD recognition is a binary classification problem, and prior works~\cite{FANCI,Distance-based,Graph_detection} utilize AGD \textit{detectors} to differentiate the AGDs from the legitimate domains. For instance, \textit{Pereira et al.}~\cite{Graph_detection} propose \textit{WordGraph} to detect dictionary-based AGDs by extracting the highly repeated domain strings.



\subsubsection{Detector knowledge collection} Adversaries seek to exploit \textit{detector} knowledge to generate DGAs with anti-detection abilities. However, security-sensitive knowledge (e.g., the model architectures and parameters) is inaccessible to adversaries, which limits the deployment of existing full-knowledge DGAs. 
Detection feedback is different from sensitive information because it is easily observable for adversaries since \textit{detectors} are obligated to inform their users of the query domain name evaluation results.
For instance, the domain name registration process usually involves a native AGD detection step, and the detection result can be acquired quickly after submission. The feedback exposes the \textit{detectors}’ vulnerabilities to users. Given that feedback, adversaries can verify the effective AGDs that can bypass the target \textit{detectors}. Thus we argue it is a promising way to exploit feedback to produce DGAs of practicality and anti-detection ability.

\vspace{-14pt}
\subsection{Motivations}
\vspace{-3pt}
We preliminarily analyze the DGAs' anti-detection ability against AGD \textit{detectors}. The evaluated DGAs include three zero-knowledge DGAs, \textit{Kraken}~\cite{Kraken1}, \textit{Gozi}~\cite{Gozi} and \textit{Suppobox}~\cite{Suppobox}, and a full-knowledge DGA, \textit{Khaos}~\cite{Khaos}.
To enrich the diversity of full-knowledge DGAs, we train \textit{Khaos} against detectors including \textit{CNN}~\cite{CNN} and \textit{LSTM}~\cite{LSTM} to generate \textit{Khaos-C} and \textit{Khaos-L}, respectively. 
For AGD \textit{detectors}, we select the state-of-the-art approaches, including the feature-based \textit{detector}: \textit{FANCI} \cite{FANCI}, neural network-based \textit{detectors}: \textit{LSTM}~\cite{LSTM} and \textit{CNN}~\cite{CNN}, graph-based \textit{detector}: \textit{WordGraph}~\cite{Graph_detection} and statistics-based \textit{detector}: \emph{statistics}~\cite{Distance-based}. 





We repetitively run each DGA 100 thousand times with different random seeds and collect the generated domains as malicious samples. Meanwhile, another 100 thousand benign samples are randomly sampled from the top 1 million websites of \textit{Alexa rank}. We then merge the domains of two groups and randomly split them into training and testing sets at a ratio of $8:2$.
The AGD \textit{detectors} are then trained on the training set and evaluated on the testing set.
AUC~\cite{AUC} is adopted to measure the detection performance because of its threshold-independent characteristic. A higher AUC represents a more accurate classification by \textit{detectors}. Moreover, we formulate the anti-detection ability of DGAs as 1-AUC, as anti-detection ability and detection ability are mutually exclusive.
The AUCs of detecting existing DGAs are presented in TABLE~\ref{tab:motivation}.

\newcounter{nodecount}
\newcommand\tabnode[1]{\addtocounter{nodecount}{1} \tikz \node  (\arabic{nodecount}) {#1};}
\tikzstyle{every picture}+=[remember picture,baseline]
\tikzstyle{every node}+=[anchor=base,
minimum width=0cm,align=center,text depth=0ex,outer sep=0pt]
\tikzstyle{every path}+=[thick, rounded corners]

\begin{table}[!htb]
\vspace{-4pt}
	\centering
	\small
	\linespread{1.0}
	\renewcommand\arraystretch{1.1} 
	\setlength{\tabcolsep}{0.6mm}
	\vspace{-10pt}
	\caption{AUCs (\%) of detecting DGAs}
	\vspace{-9pt}
	\begin{threeparttable}
		\begin{tabular}{p{0.25cm}<{\centering}p{0.9cm}<{\centering}p{0.9cm}<{\centering}p{0.9cm}<{\centering}p{0.9cm}<{\centering}p{0.9cm}<{\centering}p{0.9cm}<{\centering}}
			\ &\multicolumn{1}{c}{ } 
			&\multicolumn{3}{c}{\textbf{Zero-knowledge}}  &\multicolumn{2}{c}{\textbf{Full-knowledge}}\\ \Xcline{2-7}{0.7pt}
			\ &\multicolumn{1}{c!{\vrule width0.7pt}}{DGA}
			&\multicolumn{1}{c}{\textit{Kraken}}
			&\multicolumn{1}{c}{\textit{Gozi}}  &\multicolumn{1}{c}{\textit{Suppobox}}  &\multicolumn{1}{c}{\textit{Khaos-C}} &\multicolumn{1}{c}{\textit{Khaos-L}}        \\  \Xcline{2-7}{0.7pt}
			\multirow{5}{*}{\rotatebox{90}{\textbf{Detector}}}  
			&\multicolumn{1}{c!{\vrule width0.7pt}}{\textit{LSTM}}     &\multicolumn{1}{c}{$\textbf{98.26}~$} &\multicolumn{1}{c}{$\textbf{96.70}~$} &\multicolumn{1}{c}{$90.66~$} &\multicolumn{1}{c}{\tabnode{$\textbf{93.88}~$}} &\multicolumn{1}{c}{\tabnode{$48.54~$}} \\
			&\multicolumn{1}{c!{\vrule width0.7pt}}{\textit{CNN}}      &\multicolumn{1}{c}{$96.48~$} &\multicolumn{1}{c}{$96.20~$} &\multicolumn{1}{c}{$\textbf{97.58}~$} &\multicolumn{1}{c}{\tabnode{$53.84~$}} &\multicolumn{1}{c}{\tabnode{$\textbf{99.74}~$}} \\
			&\multicolumn{1}{c!{\vrule width0.7pt}}{\textit{FANCI}}    &\multicolumn{1}{c}{$95.54~$} &\multicolumn{1}{c}{$95.68~$} &\multicolumn{1}{c}{$85.07~$} &\multicolumn{1}{c}{\tabnode{$74.40~$}} &\multicolumn{1}{c}{\tabnode{$98.85~$}} \\ 
			&\multicolumn{1}{c!{\vrule width0.7pt}}{\textit{WordGraph}}&\multicolumn{1}{c}{$74.70~$} &\multicolumn{1}{c}{$92.56~$} &\multicolumn{1}{c}{$80.56~$} &\multicolumn{1}{c}{\tabnode{$77.82~$}} &\multicolumn{1}{c}{\tabnode{$66.34~$}} \\
			&\multicolumn{1}{c!{\vrule width0.7pt}}{\textit{Statistics}} &\multicolumn{1}{c}{$92.73~$} &\multicolumn{1}{c}{$73.87~$} &\multicolumn{1}{c}{$66.47~$} &\multicolumn{1}{c}{\tabnode{$79.41~$}} &\multicolumn{1}{c}{\tabnode{$55.94~$}}\\
			\Xcline{2-7}{0.7pt}
		\end{tabular}
		\begin{tablenotes}
			\footnotesize
			\item \textbf{Note:} For each DGA, the highest detection AUC is presented in bold. The results concerning \textit{concept drift} are in the red box.
		\end{tablenotes}
	\end{threeparttable}
	\begin{tikzpicture}[overlay]
	\node[draw=red,rounded corners = 0.25ex,fit=(1)(4),inner sep = 0pt] {};
	\node[draw opacity=0.3,rounded corners = 0.5ex,fit=(5)(10),inner sep = 0pt] {};
	\end{tikzpicture}
	\label{tab:motivation}
	\vspace{-8pt}
\end{table}

As shown in TABLE~\ref{tab:motivation}, AGD \textit{detectors} are {highly accurate} when identifying various DGAs. For example, the highest AUC can achieve up to $98.26\%$, $96.70\%$, $97.58\%$, $93.88\%$ and $99.74\%$ when detecting \textit{Kraken}, \textit{Gozi}, \textit{Suppobox}, \emph{Khaos-C} and \emph{Khaos-L}, respectively. This demonstrates the effectiveness of AGD \textit{detectors} while challenging the usability of DGAs.


\textbf{Observation \#1}: \textit{Full-knowledge DGAs present \underline{higher} \underline{ anti-detection ability} than zero-knowledge DGAs, and they decrease the detection accuracy by an average of 19.9\%.}

Zero-knowledge DGAs produce domain names relying on \textit{fixed} and \textit{detector}-independent heuristics.
The \textit{fixed} characteristic determines they generate AGDs with fixed features, e.g., the AGDs generated by \textit{Kraken} are not pronounceable~\cite{Distance-based} since they are combinations of random characters. This specific linguistic feature makes \textit{Kraken} easy to be identified, e.g., the linguistic feature-based \textit{FANCI} achieves an AUC of $95.54\%$ when detecting it.
In addition, \textit{detector}-independent characteristic determines zero-knowledge DGAs cannot adaptively update against various AGD \textit{detectors} and will become permanently ineffective once comprised by AGD \textit{detectors}.

Unlike zero-knowledge DGAs, full-knowledge DGAs produce adversarial AGDs against target \textit{detectors}.
First, the \textit{adversarial} characteristic determines the AGDs are associated with dynamical features, which vary with the targets. Such dynamical features challenge the \textit{detectors}, causing they cannot effectively detect the adversarial AGDs against various targets.
For instance, \textit{LSTM} presents AUC of $93.88\%$ when detecting \textit{Khaos-C}, while the AUC drops to $48.54\%$ when detecting \textit{Khaos-L}.
More importantly, \textit{adversarial} characteristic enables them to adaptively update models to resist different targets, e.g., \textit{Khaos} achieves the highest anti-detection ability against the \textit{CNN} and \textit{LSTM }by setting them as its targets, respectively.

\begin{figure}
	\centering
	\includegraphics[width=0.39\textwidth]{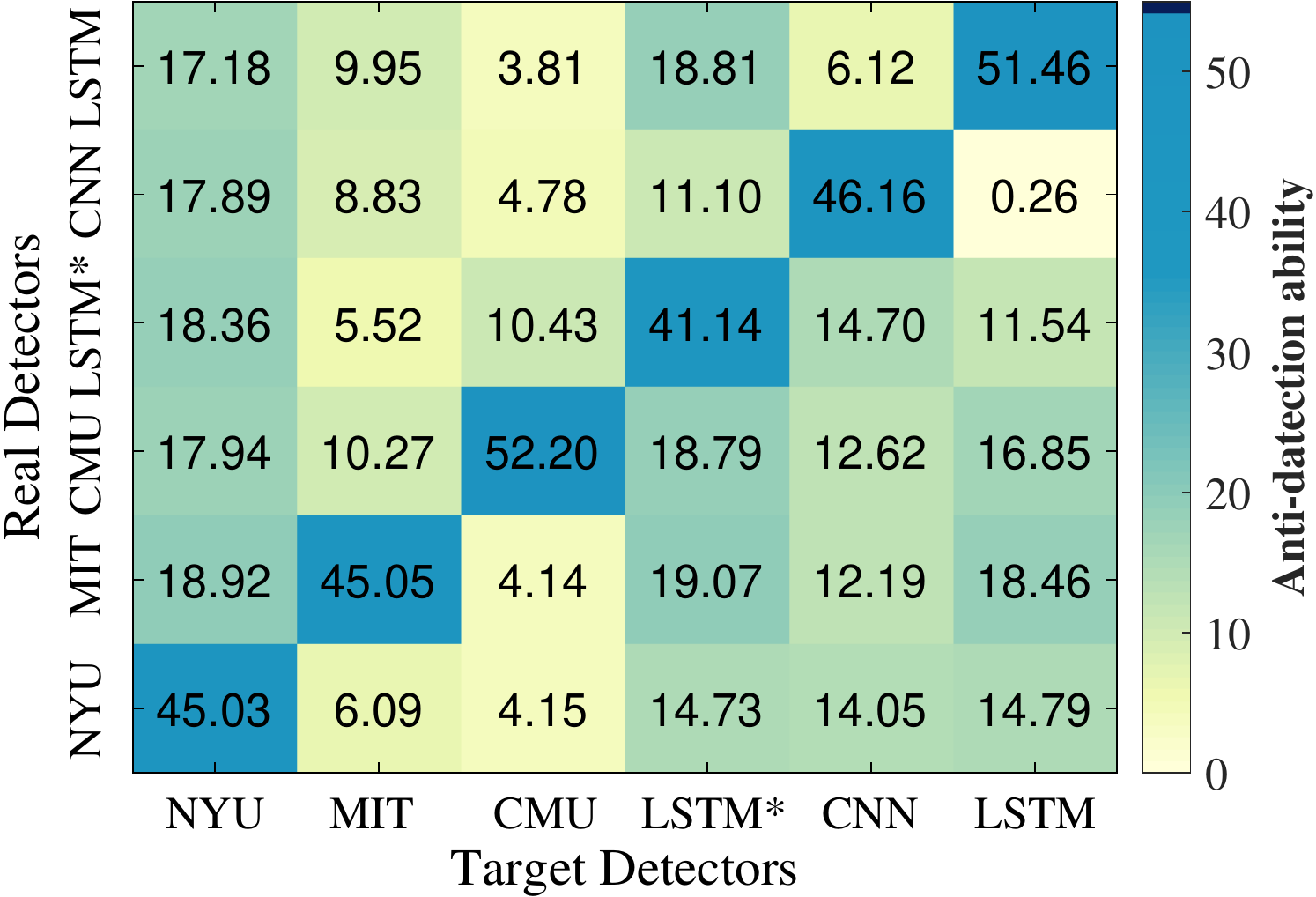}
	\vspace{-9pt}
	\caption{Target detector vs. real detector. Full knowledge DGA (\textit{Khaos}) is trained against target detector and evaluated by real detector.}
	\vspace{-20pt}
	\label{fig:motivation}
\end{figure}

\textbf{Observation \#2}: \textit{\underline{Concept drift} causes full-knowledge DGAs’ anti-detection abilities to decrease significantly.}

Concept drift refers to the anti-detection ability decrease caused by inconsistency between the targets and the real AGD \textit{detectors}. 
As shown in the red box in Table~\ref{tab:motivation}, \textit{Khaos-C} and \textit{Khaos-L} achieve $\approx$50$\%$ anti-detection ability against both the \textit{CNN} and \textit{LSTM} (where targets and real \textit{detectors} are identical). However, their anti-detection ability drops significantly once the targets and real \textit{detectors} become different. 

To further verify concept drift, we diversify \textit{Khaos} by training it against various neural-based target \textit{detectors} (including the \textit{NYU}~\cite{NYU}, \textit{MIT}~\cite{MIT}, \textit{CMU}~\cite{CMU}, \textit{LSTM*}~\cite{LSTMMI}, \textit{CNN}~\cite{CNN}, and \textit{LSTM}~\cite{LSTM}) and evaluate its anti-detection ability against those \textit{detectors}. The evaluation results are presented in Fig.~\ref{fig:motivation}, where obvious concept drifts can be observed. That is, the average anti-detection ability in the diagonal (where the targets and real \textit{detector} are the same) is $46.84\%$, much higher than the average anti-detection performance in the rest of cases. 


The reason for concept drift is analyzed as follows. Adversarial AGDs produced by \textit{Khaos} represent the vulnerabilities of the targets, that is, the target \textit{detectors} cannot identify them despite the fact that they are not benign. Different \textit{detectors} vary in terms of factors such as architecture and hyperparameters, which prevents them from sharing common vulnerabilities. Then the concept drift arises, adversarial AGDs fail to act against \textit{detectors} except for their target detectors. 
\subsection{Implications}
Existing DGAs are challenged by AGD \textit{detectors}. Although exploiting the knowledge of \textit{detectors} can help to improve the anti-detection ability of DGAs (\textbf{Observation \#1}), their native adversarial learning methods rely on the \textit{detectors} to backpropagate the gradients for model updating. 
Without knowing the \textit{detectors}, concept drift will degrade the anti-detection ability of full-knowledge DGAs (\textbf{{Observation \#2}}).

Generally, there are two solutions for addressing the problem of concept drift. First, a concept drift-robust DGA can be designed so that adversarial AGDs can maintain a high anti-detection ability even if the targets differ from the real \textit{detectors}. Second, the knowledge requirements can be reduced. The existing full-knowledge DGAs have to attack a local simulated \textit{detector} since not all of the information about the real targets is accessible. Then, concept drift arises. To this end, designing DGAs that can produce adversarial AGDs using only the easily-observed information (i.e., the feedback) is another way to counter concept drift. 
We believe the first solution is challenging since different \textit{detectors} differ in vulnerabilities. Thus, we choose the second method to mitigate concept drift.



\begin{table}
\vspace{-8pt}
 \centering 
 \setlength{\tabcolsep}{0.5mm}{
 \footnotesize
 \caption{A survey of prior DGAs }
 \vspace{-6pt}
 \label{DGA_summary}
 \begin{threeparttable}
 \begin{tabular}{p{2cm}<{\centering}l}
 \hline
 \multirow{7}{*}{Zero-knowledge} & \textit{Bamital}~\cite{Bamital},  \textit{Banjori}~\cite{Pykspa}, \textit{Bedep}~\cite{Bedep}, \textit{Conficker}~\cite{Conficker}, \\
 & \textit{Corebot}~\cite{Pykspa}, \textit{Gozi}~\cite{Gozi}, \textit{Gootkit}~\cite{Gootkit}, \textit{DirCrypt}~\cite{Pykspa},\\
 & \textit{Bazarloader}~\cite{Pykspa}, \textit{Kraken}~\cite{Kraken}, \textit{Mewsei}~\cite{Pykspa}, \textit{Hesperbot}~\cite{Hesperbot},\\
 & \textit{Suppobox}~\cite{Suppobox}, \textit{Necurs}~\cite{Pykspa}, \textit{Szribi}~\cite{Szribi}, \textit{CryptoLocker}~\cite{CryptoLocker},\\
 & \textit{Torpig}~\cite{Torpig}, \textit{UrlZone}~\cite{Pykspa}, \textit{Pykspa}~\cite{Pykspa}, \textit{GameoverP2P}~\cite{Gameover-P2P},\\
 & \textit{Murofet}~\cite{Pykspa}, \textit{Simda}~\cite{Pykspa},  \textit{charbot}~\cite{CharBot}$\ddag$, \textit{Symmi}~\cite{Pykspa}, \\
 & \textit{Geodo}~\cite{Geodo}, \textit{Ramnit}~\cite{Pykspa}, \textit{Nymaim}~\cite{Nymaim}, \textit{Zloader}~\cite{Pykspa} \\ \hline
 Full-knowledge
 & \textit{Khaos}$\ddag$~\cite{Khaos}, \textit{maskDGA}$\ddag$~\cite{MaskDGA}, \textit{DeepDGA}$\ddag$~\cite{DeepDGA} \\
 \hline
 \end{tabular}
 \begin{tablenotes}
			\footnotesize
			\item \textbf{Note:} $\ddag$ represents the DGAs have not been applied in real-life botnet.
		\end{tablenotes}
	\end{threeparttable}
 }
\vspace{-19pt}
\end{table}


Prior DGAs are either zero- or full-knowledge-based (TABLE~\ref{DGA_summary}). Assuming nothing about the \textit{detector} causes low anti-detection performance against \textit{detectors}, while the strong assumptions made by fully detector-aware algorithms hinder their real-world application (the existing knowledge-based DGAs are only utilized in  research). We argue that the \textit{feedback} that can be easily observed from the \textit{detectors} can improve the anti-detection ability and facilitate deployment in real-life scenarios. Therefore, we explore a partial-knowledge DGA for generating AGDs that can escape identification by arbitrary \textit{detectors} without requiring sensitive information.

\newcommand{\blackding}[1]{\raisebox{-1.5pt}{\large\ding{#1}}}
\vspace{-10pt}
\section{PKDGA Design}\label{sec:des}
\vspace{-4pt}
In this section, we present the design of \emph{PKDGA}.
\vspace{-14pt}
\subsection{Overview}
\begin{figure}
	\vspace{-2pt}
	\centering
	\includegraphics[width=0.48\textwidth]{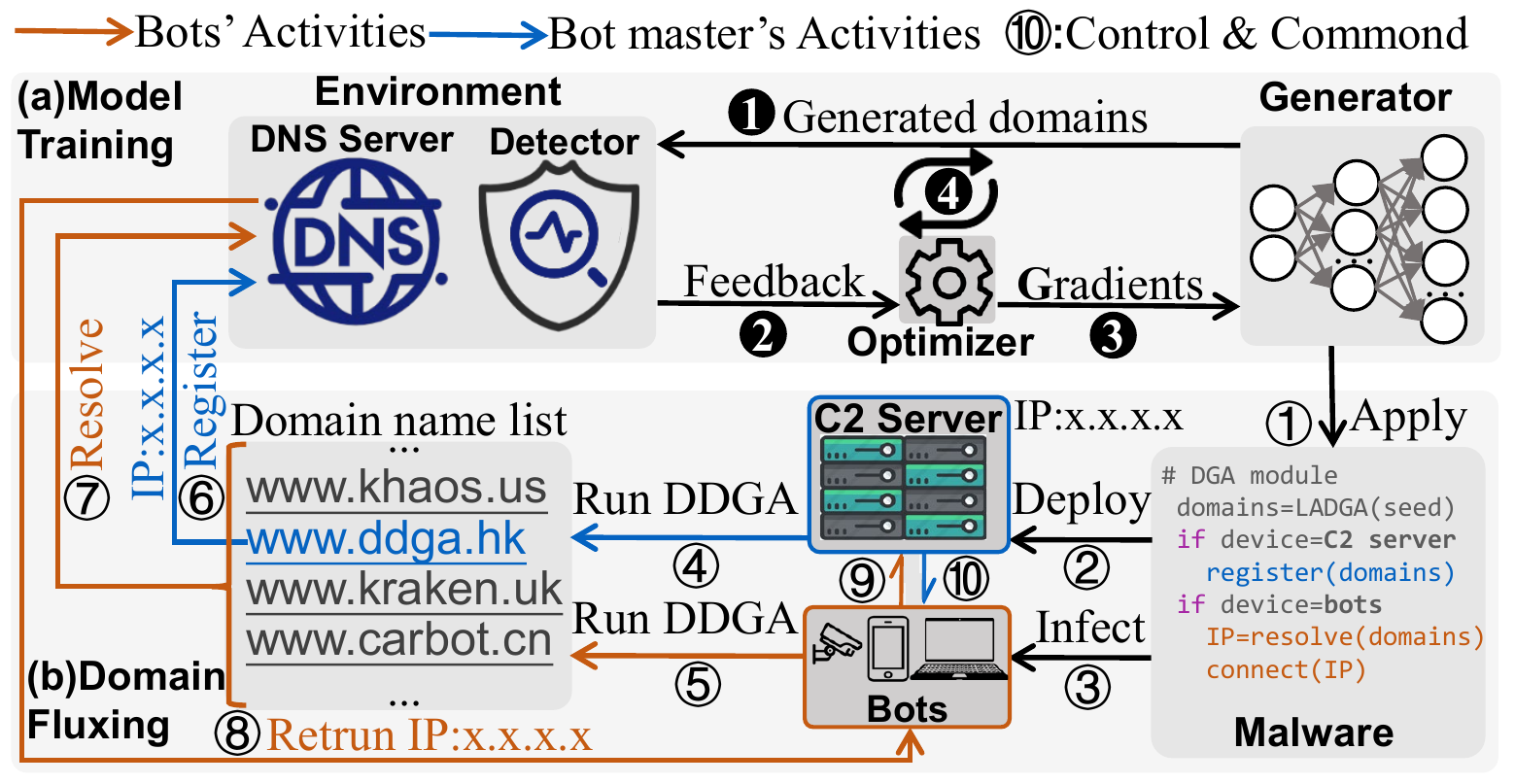}
	\vspace{-8pt}
	\caption{PKDGA Overview}
	\label{fig:overview}
	\vspace{-17pt}
\end{figure}

\emph{PKDGA} adopts the reinforcement learning (RL) architecture to explore a domain generator for maximizing the \textit{detector}-provided rewards.
RL is about the agent in the environment, learning to select the optimal action sequence using only rewards from the
environment~\cite{COMST_RL}. Thus it is well suited for generating adversarial AGDs against a black-box \textit{detector} when we consider the AGD's tokens, the feedback and the \textit{detector} as the corresponding actions, rewards and environment, respectively.  
Fig.~\ref{fig:overview} highlights the overview of \emph{PKDGA}, including two stages: \emph{model training} and \emph{domain fluxing}.

\textit{\textbf{Model training}}: The domain generator utilizes the feedback from the Domain Name System (DNS) to train the model. The training process primarily consists of four steps: \blackding{182}: The generator receives a seed and then generates a domain name. \blackding{183}: The adversary attempts to register the domain in the DNS  and derives the reward of the domain name based on the feedback from the DNS. In particular,
the generator receives a positive reward if the domain name is registered successfully, indicating that the generated domain can be utilized for domain fluxing. Otherwise, the reward is negative, implying that the queried domain is illegitimate.
\blackding{184}: Given the returned rewards, an optimizer is then activated to derive a set of gradients that are used to update the generator model. \blackding{185}\normalsize: The above steps are repeated until the generator converges.

\textit{\textbf{Domain fluxing}}: Given the learned generator, the attacker integrates it into malware and launches attacks. \whiteding{1}: The attacker firstly replaces the DGA module of a malware (e.g., a trojan or worm) with the learned AGD generator. \whiteding{2}\whiteding{3}: The adversary deploys the updated malware's botmaster program on a C\&C server and infects the victims' devices with its bot executable. \whiteding{4}\whiteding{5}: Both the C\&C server and the bots concurrently run \emph{PKDGA} with a preselected seed. As a result, they produce the same list of domain names. \whiteding{6}: The C\&C server randomly selects a domain from the candidate list and registers it with the C\&C server's IP address in DNS. In the case of registration failures, the C\&C server simply re-selects a new candidate to register until the success of registration. \whiteding{7}\whiteding{8}: The bots currently are agnostic to the registered domain names. To obtain the C\&C server's IP address, they resolve the  candidate domain names one by one until the C\&C server's address is returned. \whiteding{9}: The bots attempt to connect with the C\&C server using the resolved IP. \whitedings{10}: The C\&C server receives messages from the bots and launches attacks.

\vspace{-12pt}
\subsection{Model Training}
\vspace{-4pt}

\subsubsection{Problem description}

The problem of generating adversarial AGDs against target \textit{detector} is described as follows. 

\textit{\textbf{Generating sequential tokens}}: Domain names are sequences of tokens. We exploit a $\theta$-parameterized generator $G_\theta$ to produce token sequence $Y$ as a synthetic AGD, i.e., 
{\setlength\abovedisplayskip{1pt}
\setlength\belowdisplayskip{1.5pt}
\begin{equation}
\vspace{-3pt}
\begin{split}
Y&=[y_1,y_2,\cdots,y_T]=G_\theta(seed)\\
&s.t. \ \ \ y_{t}\in\mathbb{Y},\ 1\le t\le T, \\
\end{split}
\vspace{-3pt}
\end{equation}
where $T$ is the sequence length, and $y_{t}$ is a token sampled from the token dictionary $\mathbb{Y}$. Note that $y_t$ is the basic element of the domain names and $\mathbb{Y}$ contains all of the possible token values, including alphabetical characters (\textit{a-z} and \textit{A-Z}), digits ($0-9$) and hyphens ($\cdot$ and $/$). }

\textit{\textbf{Synthesizing domain segments}}:
A specific domain name, like ``$\operatorname{scholar.google.com}$'', can be segmented into three parts: top-level domain (``$\operatorname{com}$''), second-level domain (``$\operatorname{google}$'') and third-level domain (``$\operatorname{scholar}$''). 
It is unnecessary for a DGA to synthesize top-level domains (TLDs), since end users are not allowed to register TLDs due to the technical and operational responsibility involved in the internet's infrastructure. 
Additionally, synthesizing third-level domains (3LDs) is also meaningless. 
A 3LD can be integrated with either a legitimate second-level domain (2LD), or a malicious algorithmically-generated 2LD. The former case requires the authorization from the owner who holds the domain of 2LD, which is almost impossible. In the latter case, the AGD can always be identified through its malicious 2LD regardless of whether the 3LD is benign or not. 
Therefore, we focus only on the 2LD generation and refer to them as domain names.

\textit{\textbf{Generating adversarial domains}}: Assuming the target detector is integrated into the \textit{DNS}, the \textit{DNS} provides feedback for the query domain name $Y$ as below,
{\setlength\abovedisplayskip{1pt}
\setlength\belowdisplayskip{0.5pt}
\begin{equation}
DNS(Y)=\begin{cases}1\mbox{,} \ \ \ \mbox{if } Y   \mbox{ is successfully registered,}\\ 0 \mbox{,}  \ \ \ \mbox{if } Y  \mbox{ fail to be registered.}
\end{cases}
\label{eq:register}
\end{equation}
To produce adversarial samples that can bypass $DNS$, the objective for training $G_\theta$ is formulated as below,}

{\setlength\abovedisplayskip{-4pt}
\setlength\belowdisplayskip{-2pt}
\vspace{-12pt}
\begin{equation}
\mathop{\rm{argmax}}_{\theta}DNS(G_\theta(seed)).
\label{eq:objective}
\end{equation}
\vspace{-12pt}}


We obtain generator $G_\theta$ using RL, a technique that learns optimal actions given a certain state and is widely applied in decision-making scenarios. RL primarily consists of two components: a \textit{learner} and an \textit{explorer}. \textit{Learner} is a parameterized neural network that is trained based on rewards and objectives. \textit{Explorer} employs the trained model as a generator that accepts the current state as input and provides the actions as output. It interacts with the environment to acquire the reward for each action. Clearly, \textit{Explorer} can work without knowing the DNS’s implementation details. Therefore, RL is well suited for adversarial AGD generation based only on feedback.

\subsubsection{RL Training}

Next, we describe the RL generator’s \emph{state space}, \emph{action space}, \emph{state transition} and \emph{reward function} and show how they are used to generate AGDs. 

\textit{\textbf{State space}}:
States represent the current status of an RL task. We consider the states of the following types: (1) \emph{Initial state}, which initializes the domain generation using a seed; (2) \emph{Intermediate state}, which includes the generated token sequence so far. Formally, a state $s_t$ at time $t$ is defined as
{\setlength\abovedisplayskip{1pt}
\setlength\belowdisplayskip{1pt}
\begin{equation}
\begin{split}
&s_t\overset{def}{=}[y_0,y_1,\cdots,y_{t-1}],\\
s.t. \ \ &y_0\in \mathbb{P}\ \mbox{and} \ y_{t'\in[1,\cdots,t-1]}\in \mathbb{Y},\\
\end{split}
\end{equation}
where $y_{t'\in[1,\cdots,t-1]}$ represents the token sampled from dictionary $\mathbb{Y}$ at time $t'$, and $\mathbb{P}$ indicates the set of seeds used to start the sequence generation. We utilize date as a seed \cite{CharBot},}

{\setlength\abovedisplayskip{1pt}
\setlength\belowdisplayskip{1pt}
\vspace{-8pt}
\begin{equation}
\mathbb{P}\overset{def}{=} \{\mathcal{F}(date)|D_s\le date\le D_e\},
\end{equation}
where $\mathcal{F}(\cdot)$ refers to a function that encodes the date into a sequence with the same dimensions as $\mathbb{Y}$, and $D_{s(e)}$ represents the starting (ending) date.} 


\textit{\textbf{Action space}}: Actions are a set of operations that an RL agent can perform to change states. We consider a token to be generated as an action and define it as
{\setlength\abovedisplayskip{1pt}
\setlength\belowdisplayskip{1pt}
\begin{equation}
\mathcal{A}\overset{def}{=}\{a_1,\cdots,a_n\}, \ \ \ s.t. \ a_{i\in{[1,\cdots,n]}}\in \mathbb{Y},
\end{equation}
where $n$ refers to the size of token dictionary.}

\begin{figure}
\vspace{-12pt}
	\centering
	\includegraphics[width=0.48\textwidth]{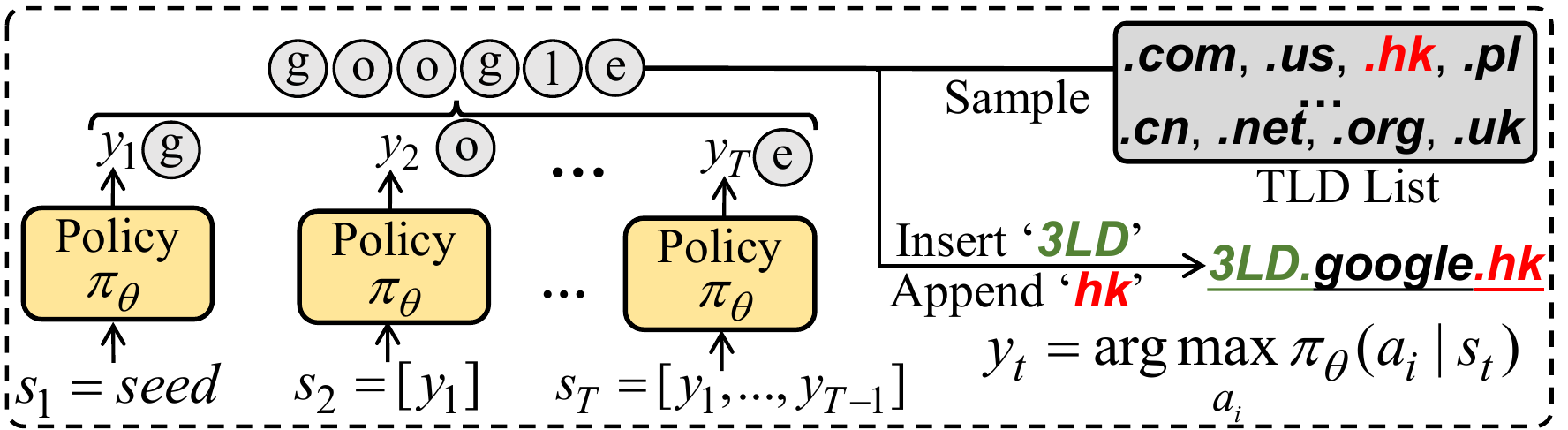}
	\vspace{-4pt}
	\caption{Generating domain names using RL. (1) At the beginning, the agent receives the starting state $s_1$, generates $y_1$ according to Eq.~\ref{eq:generation} and derives next state $s_2$ following Eq.~\ref{eq:update}. Then the agent repeats the process above until $T$ tokens are produced. (2) A valid domain name can be constructed by inserting a third-level domain (3LD) at the beginning and appending a top-level domain (TLD) at the end of synthetic sequence. Note that 3LDs are optional and randomly sampled from top 1M websites of the \textit{Alexa rank}.}
	\label{fig:domain-generation}
	\vspace{-20pt}
\end{figure}

\textit{\textbf{State transition}}: State transition refers to the change in state once an agent accepts a new action. Given the current state $s_t$ and a policy $\pi_\theta$, we derive the new token $y_t$ by selecting the action with the highest probability,
{\setlength\abovedisplayskip{1pt}
\setlength\belowdisplayskip{1pt}
\begin{equation}
y_t=\mathop{\rm{argmax}}\limits_{a_i}\pi_\theta(a_i|s_{t}),
\label{eq:generation}
\end{equation}
where $y_t$ refers to the token generated at time $t$ and $\pi_\theta(a_i|s_{t})$ represents the probability of taking action $a_t$ under state $s_t$ following policy $\pi_\theta$. Then, the next state $s_{t+1}$ is updated by concatenating the previously produced sequence $s_{t}$ with the currently generated token $y_t$. That is,}
{\setlength\abovedisplayskip{1pt}
\setlength\belowdisplayskip{2pt}
\begin{equation}
s_{t+1} = [s_{t}, y_t].
\label{eq:update}
\end{equation}
Then, we use the token sequence generated in an episode as a domain name $Y$, i.e., }
{\setlength\abovedisplayskip{1pt}
\setlength\belowdisplayskip{1pt}
\begin{equation}
Y= [y_1,y_2,\cdots,y_T].
\end{equation}
For clarity, we illustrate the process of generating token sequence using the RL method in Fig.~\ref{fig:domain-generation}. }

\textit{\textbf{Reward function}}: We define the reward of a generated domain $Y$ using the feedback from $DNS$, i.e.,
{\setlength\abovedisplayskip{1pt}
\setlength\belowdisplayskip{1pt}
\begin{equation}
R(Y) = DNS(Y).
\label{eq:reward}
\end{equation}
The reward is positive if $Y$ can be registered and vice verse.}


\begin{figure*}[!htbp]
	\vspace{-8pt}
	\centering
	\includegraphics[width=0.98\textwidth]{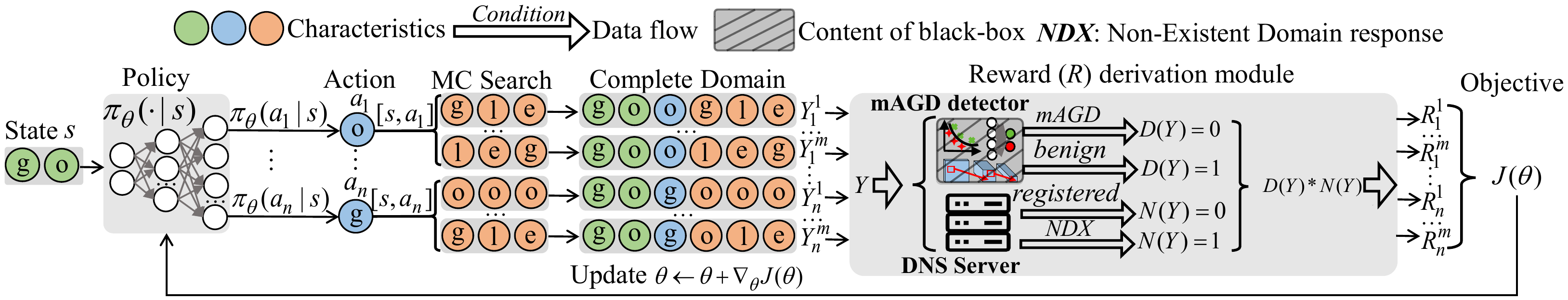}
	\vspace{-7pt}
	\caption{The detailed training process under a particular state $s$. First, the policy $\pi_\theta$ receives a particular $s$ and outputs the probabilities of taking different actions. Then, new states are derived by merging $s$ and actions. We perform MC search based on the new states and derive the complete domain name by concatenating new states with the sequence generated by MC search. Then, we evaluate the generated complete domain names and compute their rewards according to Eq.~\ref{eq:reward}. At last, we compute the gradient on $J(\theta)$ based on Eq.~\ref{eq:gradient} and update policy $\pi_\theta$ according to Eq.~\ref{eq:updating}.} 
	\label{fig:training_process}
	\vspace{-16pt}
\end{figure*}


\textit{\textbf{Policy gradient}}:
To cause the token sequence $Y$ to resist the target $DNS$, we maximize its expected reward,
{\setlength\abovedisplayskip{2pt}
\setlength\belowdisplayskip{3pt}
\begin{equation}
\begin{split}
\mathop{\rm{argmax}}_{\theta}J(\theta) &=\mathbb{E}[R(Y^{\pi_\theta}_{s_1})]\\
&=\Sigma_{i=1}^{n}\pi_\theta(a_i|s_1) R(Y^{\pi_\theta}_{[s_1,a_i]})\\
&=\Sigma_{i=1}^{n}\pi_\theta(a_i|s_1) R(Y^{\pi_\theta}_{s_2}),\\
\end{split}
\label{eq:objective_policy}
\end{equation}
where $Y^{\pi_\theta}_{s_1}$ represents a complete AGD starting from $s_1$ following policy $\pi_\theta$.
We utilize policy gradient~\cite{SeqGAN} to optimize the objective $J(\theta)$, i.e., updating $\theta$ by the gradients on $J(\theta)$,}
{\setlength\abovedisplayskip{1pt}
\setlength\belowdisplayskip{1pt}
\begin{equation}
\begin{split}
\theta = \theta+lr*\nabla_\theta J(\theta),
\label{eq:updating}
\end{split}
\end{equation}
where $lr$ is the learning rate. Like \cite{Policy-Gradient}, $\nabla_\theta J(\theta)$ is derived}
{\setlength\abovedisplayskip{5pt}
\setlength\belowdisplayskip{-9pt}
\begin{equation}
\begin{split}
\nabla_\theta J(\theta) &= \Sigma_{t=1}^{T-1}\Sigma_{i=1}^{n}\nabla_\theta\pi_\theta(a_i|s_t) R(Y^{\pi_\theta}_{[s_{t},a_i]}).\\
\label{eq:gradient}
\end{split}
\end{equation}
Using the likelihood ratio, $\nabla_\theta J(\theta)$ can be rewritten as}
{\setlength\abovedisplayskip{5pt}
\setlength\belowdisplayskip{1pt}
\begin{equation}
\begin{split}
\nabla_\theta J(\theta)&= \Sigma_{t=1}^T\Sigma_{i=1}^{n}\pi_\theta(a_i|s_t)\nabla_\theta[\log\pi_\theta(a_i|s_t) R(Y^{\pi_\theta}_{[s_{t},a_i]})]\\
& = \Sigma_{t=1}^T\mathbb{E}_{a_i\sim\pi_\theta(a_i|s_t)}\nabla_\theta[\log\pi_\theta(a_i|s_t) R(Y^{\pi_\theta}_{[s_{t},a_i]})].
\end{split}
\end{equation}
Next, we discuss how to derive $R(Y^{\pi_\theta}_{s_{t+1}})$ in two cases.}

(1) For the end state $s_T$, it can be directly derived as
{\setlength\abovedisplayskip{2pt}
\setlength\belowdisplayskip{2pt}
\begin{equation}
\begin{split}
R(Y_{[s_T,a_i]}^{\pi_\theta}) =  R([s_T, a_i]),
\end{split}
\label{eq:objective_end}
\end{equation}
where $[s_T,a_i]=[y_1,\cdots,y_{T-1},a_i]$ represents a complete token sequence by concatenating state $s_T$ and action $a_i$.}

(2) For an intermediate state $s_{t\ (1\le t<T)}$, the immediate rewards of a token sequence $[s_t, a_t]$ cannot be derived since the domain is incomplete and thus cannot be directly evaluated by the $DNS$. Instead of deriving the immediate rewards, we estimate the future reward of sequence $[s_t, a_t]$ as $R(Y_{[s_t,a_i]}^{\pi_\theta})$. In particular, given the existing tokens $[s_t, a_t]$, we estimate $m$ complete domains by sampling the remaining $T-t$ tokens using \textit{Monte Carlo} searches, i.e.,
{\setlength\abovedisplayskip{1pt}
\setlength\belowdisplayskip{2pt}
\begin{equation}
\begin{split}
[s_t,a_i, \hat{y}^{j}_{t+2}, \cdots,\hat{y}^{j}_{T}]&=MC^{\pi_\theta}([s_t,a_i];j),\\
\end{split}
\end{equation}
where $MC^{\pi_\theta}$ is the \textit{Monte Carlo} search with a roll-out policy $\pi_\theta$, and $\hat{y}^{j}_{t'}$ is the estimated token at time $t'$ in the $j$-th search.
Then, the rewards of action $a_i$ in the intermediate state is
\setlength\abovedisplayskip{2pt}
\setlength\belowdisplayskip{2pt}
\begin{equation}
R(Y_{[s_t,a_i]}^{\pi_\theta})=\Sigma_{j=1}^m R(MC^{\pi_\theta}([s_t,a_i];j))/m,
\label{eq:objective_intermediate}
\end{equation}
where $m$ is a hyper-parameters, representing the number of MC searches. For clarity, we describe the detailed process of training RL policy to produce adversarial domains in Fig.~\ref{fig:training_process}. }

We choose the \emph{stochastic gradient descent} (SGD) method to optimize the policy gradient-based objective (Eq.~\ref{eq:objective_policy}). Before starting an SGD optimizer, we need to initialize two key hyperparameters: the learning rate $lr$ and the batch size $b$, which likely significantly impact the training performance and efficiency. 
Therefore, we utilize a grid search to find the optimal values of them (shown in TABLE~\ref{tab:HP-tuning}). 

\vspace{-12pt}
\subsection{Training Policy} 
\vspace{-4pt}
The training policy $\pi_\theta$ actually belongs to a natural language processing (NLP) task, i.e., predicting the next token of the maximal reward based on the given/observed token sequence. As the Recurrent Neural Network (\textit{RNN}) has shown great success in NLP~\cite{deeplog,inter_speech}, we adopt the \textit{RNN} model as the architecture of policy $\pi_\theta$. In particular, we choose the \textit{LSTM} network because \textit{LSTM} can avoid the gradient explosion or gradient vanishing caused by the long token sequences~\cite{LSTM_pro}. 

\begin{figure}[!htbp]
	\vspace{-8pt}
	\centering
	\includegraphics[width=0.42\textwidth]{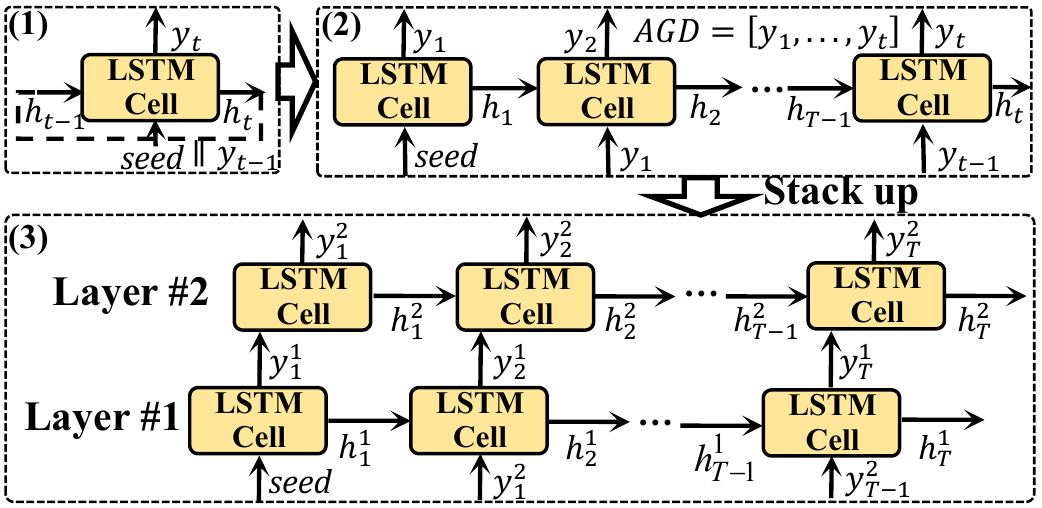}
	\vspace{-8pt}
	\caption{LSTM architecture illustration. \textbf{(1)}: The recurrent characteristic of LSTM; \textbf{(2)}: The unfolded LSTM; \textbf{(3)}: The stacked LSTM. }
	\label{fig:LSTM}
	\vspace{-16pt}
\end{figure}

Fig.~\ref{fig:LSTM} illustrates the architecture of the \textit{LSTM} network, where the \textit{LSTM cell} is the basic processing unit and constructed by a set of gating functions~\cite{LSTM_intro}. The \textit{LSTM} network adopts a recurrent structure, i.e., the outputs of the last step are forwarded as the inputs of the next step. 
At a particular time step $t$, each \textit{LSTM cell} reads an external input (i.e,. seed or $y_{t-1}$ the token generated in the last step) and the hidden state $h_{t-1}$ of its parent step. Then, each \textit{LSTM cell} processes those inputs and produces an output $y_t$ and a hidden state vector $h_t$.

The dimension of $y_t$ (denoted by $d_y$), the dimension of $h_t$ (denoted by $d_h$) and the dimension of input embedding (denoted by $d_e$) are hyper-parameters and need to be given prior to building \textit{LSTM} network.
In our case, $y_{t}$ is the probability of taking different tokens, and thus $d_y$ is identical to the token dictionary size. Moreover, $d_h$ and $d_e$ determine how much of sequential information and input information are maintained. Stacked \textit{LSTM} network can be derived by stacking up \textit{LSTM cells} and forwarding the output of the previous layer as the input of each corresponding \textit{LSTM cell} in the next layer (part (3) in Fig.~\ref{fig:LSTM}), where the number of stacked \textit{LSTM} layers (denoted by $N_l$) is a hyper-parameter. We tune the hyper-parameters $d_h$, $d_e$ and $N_l$ by grid search.  

\vspace{-4pt}
\section{Implementation}
\label{sec:imp}
We have implemented \emph{PKDGA} as a domain generation tool and released it publicly~\cite{PKDGA}. The tool provides interfaces that enable users to initialize, start training, and generate the domain names without knowing the design details. 
To use it, users need to first initialize a domain generator using \emph{generator=PKDGA(\textit{policy\_net})}, where $\operatorname{\textit{policy\_net}}$  specifies the network architecture of the policy agent. After initialization, users can train the generator using \emph{generator.train(\textit{detector\_target}, \textit{num\_MC}, \textit{opt}, \textit{epoch})}, where \emph{\textit{detector\_target}} specifies the target AGD \textit{detector}, and $\operatorname{\textit{mun\_MC}}$ refers to the number of Monte Carlo searches. Moreover, $\operatorname{\textit{opt}}$ represents the optimizer to maximize \emph{PKDGA}'s rewards. To configure an optimizer, users are required to specify the training-related hyperparameters, including the optimization method (e.g., SGD), learning rate, batch size and the parameter set they need to tune. After model training, users can generate domain names using \emph{generator.inference(\textit{seeds})}, where $\operatorname{\textit{seeds}}$ is the seed collection.

Moreover, we built a domain fluxing testbed by implementing several components, including infected devices (bots), a C\&C server, a DNS server and AGD \textit{detectors}. The interactions between the different components are shown in Fig.~\ref{fig:overview}.

\textit{\textbf{Bots}}: The bots refer to the malware-infected machines. Since we focus only on the domain generation process, we craft bot by running malware on machines without infection processes. Among the available malwares, we select \textit{Mirai} as it is the most widely spread malware since Aug. 2016 \cite{Mirai}.  

\textit{\textbf{C\&C server}}: We build a C\&C server by running the server executable of \textit{Mirai} on an individual machine. It generates a collection of candidate domain names, randomly selects one domain name from the candidates and registers that domain. 

\textit{\textbf{DNS server}}: Instead of exploiting publicly available DNS services, we establish a private DNS server since we need to specify AGD \textit{detectors} for DNS server to evaluate \textit{PKDGA}'s performance against different targets. It is nearly impossible to achieve using the public DNS services. The lightweight Linux tool \textit{dnsmasp}~\cite{DNS} is utilized to build our private DNS server. Moreover, we simulate the feedback of DNS server as 
{\setlength\abovedisplayskip{1pt}
\setlength\belowdisplayskip{1pt}
\begin{equation}
DNS(Y) = D(Y)*N(Y),
\label{eq:reward1}
\end{equation}
where $D(Y)$ is the evaluation of $Y$ scored by \textit{detector} $D$,}
{\setlength\abovedisplayskip{1pt}
\setlength\belowdisplayskip{2pt}
\begin{equation}
\begin{split}
D(Y) = \begin{cases}1\mbox{,} \ \ \  \ \mbox{if\ } Y \mbox{\ legitimate,}\\ 0 \mbox{,}  \ \ \ \ \mbox{if } Y  \mbox{ is AGD.}
\end{cases}
\end{split}
\label{eq:register1}
\end{equation}
The term $N(Y)$ measures the novelty of $Y$. That is, $N(Y)$=$1$ represents $Y$ is novel, while $N(Y)$=$0$ indicates $Y$ is registered.} 
The rationale of $DNS(Y)$ is two-fold. First, a registrable domain should be able to bypass the AGD \textit{detector} integrated into DNS. Moreover, the domain should be novel since registering a domain that has already been registered is not allowed.

\vspace{-8pt}
\section{Evaluation}\label{sec:eva}
\vspace{-2pt}
In this section, we present our evaluation of \emph{PKDGA}.
\vspace{-12pt}

\subsection{Methodology}\label{Sec:dataset}
\vspace{-2pt}
\textit{\textbf{Datasets}}: 
We build a training dataset consisting of 100 thousand benign domains and 100 thousand AGDs. In particular, the 100 thousand benign domains are randomly sampled from the top 1M sites of the \textit{Alexa rank}~\cite{Alexa}. The 100 thousand AGDs are randomly sampled from \textit{DGArchive}~\cite{DGArchive}, which records over 100 million malicious domains used by 93 mainstream malware families. 
We further build a separate testing dataset for each DGA. The AGDs are collected by solely running each DGA 100 thousand times with different seeds. 
By default, the AGD \textit{detectors} are trained on the training dataset and evaluated on the testing dataset.


\textit{\textbf{Competing DGAs}}:
Table~\ref{DGA_summary} summarizes the previously developed DGAs. We compare \emph{PKDGA} with \emph{Khaos}, \emph{Kraken}, \emph{Gozi} and \emph{Suppobox} because they are state-of-the-art approaches and cover both zero- and full-knowledge DGAs.

\begin{itemize}
\item \emph{Khaos}~\cite{Khaos} trains a \textit{Wasserstein generative adversarial network} (WGAN)~\cite{WGAN} to synthesize AGDs. According to \textit{Yun et al.}~\cite{Khaos}, we set the embedding size as $5000$. Moreover, the learning rate and the batch size are set as $10^{-3}$ and $64$, which are tuned via grid-search method. 
\item  \emph{Kraken}~\cite{Kraken1} is used by the \textit{Kraken} malware family. It uses a linear congruential generator as a pseudo-random generator to yield characters and then merges those characters in their generation order to construct domains. 

\item \emph{Gozi}~\cite{Gozi} is DGA used by the \textit{Gozi} malware family. It constructs domain names by generating words selected by the pseudo-random algorithm (the same as \textit{Kraken}) and concatenates those words as a domain name. 

\item \emph{Suppobox}~\cite{Suppobox} is the DGA used by the \textit{Suppobox} malware family, which maintains two dictionaries of commonly used English words. When generating AGDs, \textit{Suppobox} firstly samples word from each dictionary and subsequently merge the selected words as an AGD.
\end{itemize}




\textit{\textbf{AGD detectors}}:
To evaluate the anti-detection ability of DGAs, we also implement the following AGD \textit{detectors}.
\begin{itemize}

\item \emph{Statistics}~\cite{Distance-based} determines a domain as benign if it has a similar distribution with the legitimate domains. Specifically, the distribution metrics it uses are \textit{Kullback-Leibler} divergence, \textit{Jaccard} index and \textit{Edit} distance, respectively.

\item \emph{WordGraph}~\cite{Graph_detection} constructs a graph describing the relationships between 
\textit{the largest common substrings} in a domain, and the graph with average degrees larger than a predefined threshold is considered as malicious. In our implementation, the substrings that repeat more than three times are considered common substrings.

\item \emph{FANCI}~\cite{FANCI} is able to classify non-existent domains (NXDs) into DGA-generated ones and benign ones. 
It extracts 21 lightweight DGA features, including structural, linguistic and statistical features, and employs a supervised model (\textit{random forest}) for classification. 

\item \emph{NN} \textit{detectors}~\cite{LSTM,CNN} directly leverage neural networks to distinguish AGDs from benign ones without extracting features manually. We select three representative neural \textit{detectors}: \emph{CNN}~\cite{CNN}, \emph{LSTM}~\cite{LSTM} and \emph{Bi-LSTM}~\cite{LSTM}. 

\end{itemize}

\textit{\textbf{Metrics}}: We choose the threshold-independent AUC as the evaluation metric, which is derived using both  the \emph{true positive rate} (\textit{TPR}) and the \emph{false positive rate} (\textit{FPR}). Given the ground-truth and a prediction result, we can derive the \emph{true positive} (\textit{TP}), \emph{false positive} (\textit{FP}), \emph{true negative} (\textit{TN}) and \emph{false negative} (\textit{FN}) metrics.
Then the \emph{TPR} and \emph{FPR} are derived as
{\setlength\abovedisplayskip{2pt}
\setlength\belowdisplayskip{1pt}
\begin{equation}
\begin{split}
TPR &= TP/(TP+FP), \\
FPR &= FP/(FP+TN).
\end{split}
\end{equation}
As both the values of \emph{TPR} and \emph{FPR} vary across the decision thresholds, plotting \emph{(FPR, TPR)} yields a \emph{receiver operating characteristic} (ROC) curve. The area under the ROC curve is defined as the AUC. From the above processes, we can find that AUC can avoid the hassle of searching an empirical threshold for classier by computing the overall performance under all possible threshold settings.}

\vspace{-12pt}
\subsection{Hyper-parameter Tuning} 
Since the capability of deep learning is typically impacted by hyper-parameters (HPs), an effective HP tuning solution is necessary to maximize \emph{PKDGA}'s performance. In particular, there are three HP types when building and training PKDGA.
(1) \emph{Architecture-related HPs} refer to the parameters that determine the structure of \textit{LSTM cells}, including the number of layers $N_l$, the dimension of hidden state $d_h$ and the dimension of input embedding $d_e$. To find the optimal combination of HPs, we explore $N_l$, $d_h$ and $d_e$ in the range of $[1,2]$, $[2^1,2^2,\cdots,2^8]$ and $[2^1,2^2,\cdots,2^8]$, respectively.
(2) \emph{RL-related HPs} refer to the number of \textit{Monte Carlo} searches (denoted as $m$). The possible values of $m$ are selected from the set of $[10,15,20]$.
(3) \emph{Training-related HPs} include the learning rate $lr$ and batch size $b$, which are tested using $[10^{-1}, 10^{-2}, 10^{-3}, 10^{-4}]$ and $[2^1,2^2,\cdots,2^{10}]$, respectively.
Following the prior works~\cite{FlowPrint, TNSM}, a grid search is exploited to tune \textit{PKDGA}'s hyper-parameters, i.e., searching the optimal HPs iteratively. The specific processes of grid search are described as follows. (1) Set all HPs as their default values and select a HP. (2) Train \emph{PKDGA} against the target AGD \textit{detector} under all possible values of the current HP. (3) Set the value by which the maximal reward is achieved as the new default value for the current HP. (4) Select the next HP. Repeat steps (2)-(4) until all HPs are well-tuned.  The detailed iterations of tuning \emph{PKDGA}'s HPs are presented in TABLE~\ref{tab:HP-tuning} (the optimal HP of each iteration is in bold).

\begin{table}[!htb]
	\centering
	\footnotesize
	\linespread{1.5}
	\renewcommand\arraystretch{1.1} 
	\setlength{\tabcolsep}{0.6mm}
	\vspace{-10pt}
	\caption{Hyperparameter tuning process}
	\vspace{-8pt}
	\begin{tabular}{p{0.15cm}<{\centering}p{0.75cm}<{\centering}p{0.70cm}<{\centering}p{0.70cm}<{\centering}p{0.70cm}<{\centering}p{1.4cm}<{\centering}p{0.95cm}<{\centering}cp{1.35cm}<{\centering}}\Xhline{0.9pt}
		\ &\multicolumn{1}{c}{ } &\multicolumn{3}{c}{Architecture-related}  &\multicolumn{1}{c}{RL-related} &\multicolumn{2}{c}{Training-related} &Reward \\ \Xhline{0.7pt} 
		\multicolumn{2}{c!}{HP}  &\multicolumn{1}{c}{$N_{l}$}  &\multicolumn{1}{c}{$d_{e}$}    &\multicolumn{1}{c!}{$d_{h}$}    &\multicolumn{1}{c!}{$m$}  &\multicolumn{1}{c}{$lr$} &\multicolumn{1}{c!}{$b$} &$--$ \\ \Xhline{0.7pt}  
		\multicolumn{2}{c!}{Default}
		&\multicolumn{1}{c}{$2$}  &\multicolumn{1}{c}{$16$}  &\multicolumn{1}{c!}{$16$}  &\multicolumn{1}{c!}{$15$}  &\multicolumn{1}{c}{$0.001$}  &\multicolumn{1}{c!}{$32$}     &$0.9217$       \\
		\multirow{7}{*}{\rotatebox{90}{\ \ \ \ Iterations}} 	
		&\multicolumn{1}{c}{\#1}  &$\textbf{1}$  &$16$  &$16$  &$15$  &$0.001$  &$32$     &$0.9280$   \\ 
		&\multicolumn{1}{c}{\#2}  &$1$  &$\textbf{32}$  &$16$  &$15$  &$0.001$  &$32$     &$0.9286$   \\
		&\multicolumn{1}{c}{\#3}  &$1$  &$32$  &$\textbf{64}$  &$15$  &$0.001$  &$32$     &$0.9313$   \\
		&\multicolumn{1}{c}{\#4}  &$1$  &$32$  &$64$  &$\textbf{20}$  &$0.001$  &$32$     &$0.9429$   \\
		&\multicolumn{1}{c}{\#6}  &$1$  &$32$  &$64$  &$20$ &$\textbf{0.001}$  &$32$     &$0.9527$   \\
		&\multicolumn{1}{c}{\#7}  &$1$  &$32$  &$64$  &$20$  &$0.001$  &$\textbf{64}$     &$0.9885$   \\
		\Xhline{0.7pt}
	\end{tabular}
	\label{tab:HP-tuning}
	\vspace{-16pt}
\end{table}

\subsection{Anti-detection Ability}
\vspace{-1pt}
The DGAs' anti-detection ability is closely related to the detection performance of the target \textit{detectors}, that is, higher detection performance corresponds to lower anti-detection ability.  
Furthermore, the targets' detection performance is determined by AGD \textit{detectors} and training sets. To this end, we evaluate \textit{PKDGA}'s anti-detection ability by varying the AGD \textit{detector} types and the training sets, respectively.

\subsubsection{Anti-detection ability vs. AGD detectors}

We train various AGD \textit{detectors} on the unified training set (already described in Sec~\ref{Sec:dataset}) and evaluate DGAs' anti-detection ability against those \textit{detectors} given different \textit{detector} knowledge.
When the required information about targets is not available, we train \textit{Khaos} and \textit{PKDGA} against the \textit{CNN} and \textit{LSTM} by default. The experimental results are shown in TABLE~\ref{tab:anti-detection}.


\textit{\textbf{Zero-knowledge case}}: \underline{\textit{PKDGA presents comparable per-}} \underline{\textit{formance with the competing DGAs}}.
In particular, the average anti-detection ability of \emph{Kraken}, \emph{Gozi}, \emph{Suppobox}, \emph{Khaos} and \emph{PKDGA} is $8.29\%$, $22.87\%$, $23.47\%$, $23.01\%$ and $26.53\%$, respectively. 
The reason that knowledge-based DGAs do not exhibit the expected higher anti-detection performance is analyzed as follows. When target \textit{detector} information is unavailable, knowledge-based DGAs (\textit{Khaos} and \textit{PKDGA}) have to act against the simulated \textit{detector}. Then, they work in the same manner as the zero-knowledge DGAs, producing the target \textit{detector}-independent AGDs and failing to realize their advantages in terms of anti-detection ability.
Moreover, we find that \emph{Kraken} demonstrates the lowest anti-detection ability against all of the AGD \textit{detectors}. This is because \emph{Kraken} generates AGDs by concatenating pseudo-randomly sampled characters instead of words. Those character-based AGDs are obviously different from benign ones, composed of readable syllables and acronyms~\cite{Khaos} and easily detected.

\begin{table}
\centering
\small
\linespread{1.2}
\renewcommand\arraystretch{1.1} 
\setlength{\tabcolsep}{1mm}
\caption{Anti-detection ability (\%) vs. AGD detectors}
\vspace{-8pt}
\begin{threeparttable}
\begin{tabular}{p{0.2cm}<{\centering}p{1.1cm}<{\centering}p{1.1cm}<{\centering}p{1.0cm}<{\centering}p{1.0cm}<{\centering}p{1.0cm}<{\centering}p{1.0cm}<{\centering}p{0.9cm}<{\centering}}
\Xcline{2-8}{0.7pt}
 &\textbf{DGAs} & \textbf{\textit{Statistics}} & \textbf{\textit{WordG.}} & \textbf{\textit{FANCI}} & \textbf{\textit{LSTM}} & \textbf{\textit{BiLSTM}} & \textbf{\textit{CNN}} \\ 
\Xcline{2-8}{0.7pt} \specialrule{0em}{1.5pt}{1.5pt} \Xcline{2-8}{0.7pt}
\multirow{5}{*}{\rotatebox{90}{\scriptsize{\textbf{Zero Knowledge}}}} 
& \textit{Kraken}   & 6.58	& 19.69	& 9.16	& 4.14	& 3.40	& 6.79 \\
& \textit{Gozi}     & 26.72	& 7.95	& 30.36	& 17.44	& 26.62 &	28.10 \\
& \textit{Suppobox} & 33.47	& 19.58	& 36.29	& 11.79	& 15.68 &	23.98 \\
& \textit{Khaos}    & 20.59* & 22.18* & 	25.60* &	11.13* & 	6.12* & 	\textbf{52.44} \\  
& \textit{PKDGA}    & \textbf{24.12}*	& \textbf{20.88}*	& \textbf{19.74}*	& \textbf{49.57}	& \textbf{21.13}* &	23.77* \\\Xcline{2-8}{0.7pt}
\multirow{5}{*}{\rotatebox{90}{\scriptsize{\textbf{Partial Knowledge}}}}
& \textit{Kraken}   & 6.58	& 19.69	& 9.16	& 4.14	& 3.40	& 6.79 \\
& \textit{Gozi}     & 26.72	& 7.95	& 30.36	& 17.44	& 26.62 &	28.10 \\
& \textit{Suppobox} & 33.47	& 19.58	& 36.29	& 11.79	& 15.68 &	23.98 \\  
& \textit{Khaos}    & 20.59* & 22.18* & 	25.60* &	11.13* & 	6.12* & 	\textbf{52.44} \\ 
& \textit{PKDGA}    & \textbf{47.21}	& \textbf{44.51}	& \textbf{45.91}	& \textbf{49.57}	& \textbf{48.70} &	49.30 \\\Xcline{2-8}{0.7pt}
\multirow{5}{*}{\rotatebox{90}{\scriptsize{\textbf{Full Knowledge   }}}}
& \textit{Kraken}   & 6.58	& 19.69	& 9.16	& 4.14	& 3.40	& 6.79 \\
& \textit{Gozi}     & 26.72	& 7.95	& 30.36	& 17.44	& 26.62 &	28.10 \\
& \textit{Suppobox} & 33.47	& 19.58	& 36.29	& 11.79	& 15.68 &	23.98 \\  
& \textit{Khaos}    & 20.59* & 22.18* & 25.60* & 47.40 & \textbf{52.20} & \textbf{52.44}\\
&\textit{PKDGA}    & \textbf{47.21}	& \textbf{44.51}	& \textbf{45.91}	& \textbf{49.57}	& 48.70 &	49.30 \\ \Xcline{2-8}{0.7pt}
\end{tabular}
\begin{tablenotes}
	\footnotesize
	\item \textbf{Note:} Symbol \textbf{*} represents that concept drifts arise due to lack the required knowledge. The highest anti-detection performance against each AGD detector is presented in bold.   
\end{tablenotes}
\label{tab:anti-detection}
\end{threeparttable}
\vspace{-20pt}
\end{table}

\textit{\textbf{Partial knowledge case}}: \textit{\underline{PKDGA achieves significant im-} \underline{provements over the competing DGAs.}}
Specifically, compared with \emph{Khaos}, \emph{Kraken}, \emph{Gozi} and \emph{Suppobox}, \emph{PKDGA} improves the average anti-detection ability by $24.52\%$, $39.24\%$, $24.67\%$ and $24.07\%$, respectively.
Moreover, \textit{PKDGA}'s average anti-detection performance approaches 50\%, i.e., the corresponding detection AUC is approximately 50\%, which indicates that \textit{detectors} can not distinguish AGDs from benign domains. 


Two reasons account for \textit{PKDGA}'s superior anti-detection performance.
First, interacting with target \textit{detectors} using feedback enables \textit{PKDGA} to implicitly explore the vulnerabilities of targets and further generate adversarial AGDs compromising the targets.
Therefore, it outperforms the zero-knowledge DGAs (\emph{Kraken}, \emph{Gozi} and \emph{Suppobox}), which generate \textit{detector}-independent AGDs without utilizing any target information. 
Second, generating AGDs merely using feedback enables \textit{PKDGA} to avoid the concept drift caused by the lack of full information. Thus, it outperforms \textit{Khaos}, which works relying on access to all of the target’s information and suffers from concept drift when only partial knowledge is provided. 
\begin{figure*}[!htb]
	\vspace{-8pt}
	\centering
	\includegraphics[width=0.98\textwidth]{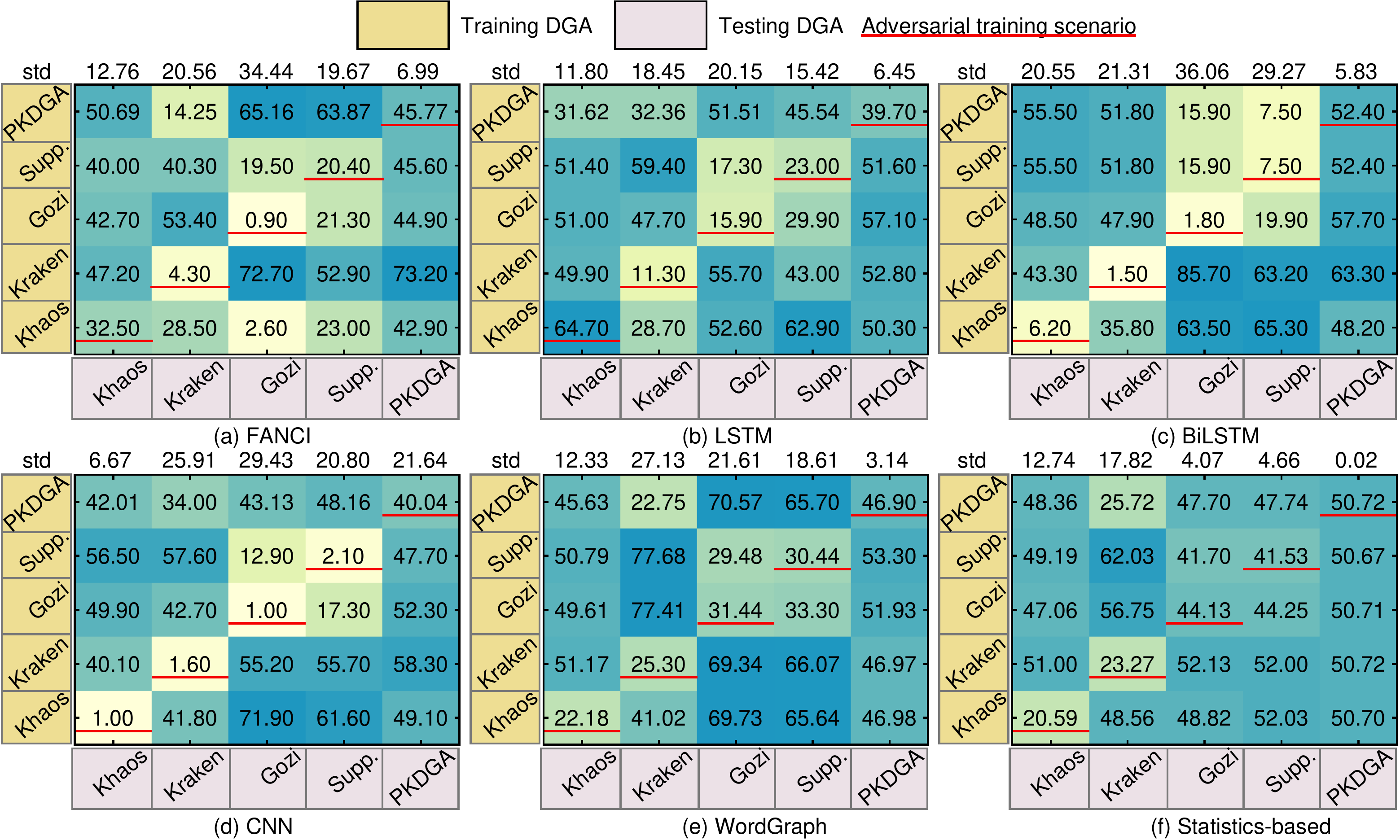}
	\vspace{-8pt}
	\caption{Anti-detection ability (\%) vs. training sets. Each AGD \textit{detector} is trained using the DGA on the Y-axis and evaluated using the DGA on the X-axis.}
	\label{fig:AUC_DGAs}
	\vspace{-18pt}
\end{figure*}

\textit{\textbf{Full-knowledge case}}: \textit{\underline{PKDGA shows overall higher perfor}-\underline{mance than the full-knowledge DGAs}}. \textit{PKDGA}'s average anti-detection ability against \textit{NN} \textit{detectors} (\textit{LSTM}, \textit{BiLSTM} and \textit{CNN}) is 49.19\%, which is slightly lower than \textit{Khaos} (50.68\%).
The performance gap between \textit{PKDGA} and \textit{Khaos} is caused by their different learning paradigm. 
In particular, \textit{Khaos} exploits \textit{adversarial learning} to solve an adversarial AGD, and its solution is guaranteed to be optimal among all of the SGD optimizer-based solutions. If not, the SGD optimizer will continue to tune the AGD generator until it becomes optimal. 
\textit{PKDGA} uses \textit{reinforcement learning} to produce an adversarial AGD, in which an exploration method is involved (i.e., \textit{Monte Carlo} search). While the optimal solution may be missed during the exploration processes, thus AGD solved by reinforcement learning can not be guaranteed to be optimal. This limitation dictates that \textit{PKDGA} is not be more effective than \textit{Khaos} for \textit{NN}-based targets. However, the performance gap between \textit{PKDGA} and \textit{Khaos} is minute (only 1.49\%), which demonstrates that \textit{Khaos} can solve an approximation of the optimal solution.  

Despite its slightly inferior to \textit{Khaos} for \textit{NN}-based targets, \textit{PKDGA} achieves much higher anti-detection performance than \textit{Khaos} when targets are not based on \textit{NN} architectures.
This is because \textit{Khaos} fails to act against \textit{non-NN} \textit{detectors} even if full knowledge is provided. In particular, \textit{Khaos} updates its AGD generator relying on the targets to compute gradients, which is impossible for \textit{non-NN} models. 
Then \textit{Khaos} must be trained against the default target, causing concept drifts when evaluating the performance.  
Different from \textit{Khaos}, \textit{PKDGA} updates model only relying on the targets' feedback, which can be provided by \textit{detectors} of any type. Thus it can avoid concept drift and shows higher performance than \textit{Khaos}. 


In the following performance evaluations, we release \textit{detectors}' partial-knowledge (feedback) to DGAs by default for the following reasons. 
First, releasing partial knowledge is a more practical setting than giving full-knowledge since the sensitive information is inaccessible in practice. 
Additionally, by using the observable partial knowledge, \textit{PKDGA} has higher potential than in the case in which it does not utilize any knowledge.
\subsubsection{Anti-detection ability vs. training sets}

To evaluate DGAs' anti-detection ability on different training datasets, we further generate five individual training sets using five DGAs. In each training set, the AGD samples are collected by solely running an individual DGA 100 thousand times, and the benign samples are sampled from the \textit{Alex rank}. We train AGD \textit{detectors} on the individual training set and evaluate the DGAs' anti-detection ability against those \textit{detectors} whose partial knowledge is given. 
The evaluation results are presented in Fig.~\ref{fig:AUC_DGAs}, where the \textit{training DGAs} are utilized to construct the individual training sets, and the \textit{testing DGAs} are evaluated in terms of their anti-detection ability. 

\textit{\textbf{Finding \#1:}} \underline{\textit{PKDGA is robust to training sets.}}
Fig.~\ref{fig:AUC_DGAs} shows that regardless of which dataset is used to train the \textit{detector}, \emph{PKDGA's} anti-detection performance is much more stable and generally higher than others.
In practice, as the training sets of AGD \textit{detectors} are usually not released publicly, the anti-detection ability of DGAs could be easily compromised by \textit{detectors}. Using Fig.~\ref{fig:AUC_DGAs}(a) as an example, \textit{Gozi}'s anti-detection ability against \textit{FANCI} is 72.7\% when \textit{FANCI} is trained on the training dataset generated by \textit{Kraken}, indicating it can compromise \textit{FANCI} in this case. However, its anti-detection performance drops to 2.60\% when \textit{FANCI} is trained using the training dataset generated by \textit{Khaos}. In contrast, \emph{PKDGA} is not sensitive to training sets and can keep good anti-detection ability ($\ge39.7\%$) on different sets.

\vspace{-8pt}
\textit{\textbf{Finding \#2:}} \underline{\textit{PKDGA is able to resist adversarial training.}}
\emph{PKDGA} is a type of \textit{adversarial example attack}~\cite{adversarial_example}; these attacks craft malicious inputs (the so-called adversarial examples) to compromise the accuracy of classifiers. Furthermore, adversarial training is introduced as an effective countermeasure against adversarial example attacks by training classifiers using adversarial examples~\cite{AWA}. 
To this end, we analyze the performance of \emph{PKDGA} in the adversarial training scenario, i.e., the scenario in which the training DGA and the testing DGA are identical. Fig.~\ref{fig:AUC_DGAs} shows that \textit{PKDGA} demonstrates much higher anti-detection ability than others under adversarial training cases, which indicates that \emph{PKDGA} can still bypass the \textit{detectors} even when they are trained using the AGDs produced by itself. 
However, the competing DGAs show undesirable performances against adversarial training, e.g., \textit{Suppobox}'s anti-detection ability against various \textit{detectors} in adversarial training cases is as low as $2.10\%$. 
This \textit{adversarial characteristic} enables \textit{PKDGA}'s resistance against adversarial training strategies. That is, even if the \textit{detectors} can identify the AGDs generated by \textit{PKDGA} via  
adversarial training, \textit{PKDGA} can still reproduce novel adversarial AGDs using the feedback and further compromise the target \textit{detector}.
The resistance against adversarial training is of great importance. Specifically, a security vendor can imitate the target DGAs by decompiling bot executable and train its \textit{detector} using the adversarial AGDs. In that case, \emph{PKDGA} can still compromise the \textit{detector} by generating novel adversarial domains.

\vspace{-12pt}
\subsection{Overheads}
\vspace{-2pt}

\begin{figure}
	\vspace{-8pt}
	\centering
	\includegraphics[width=0.45\textwidth]{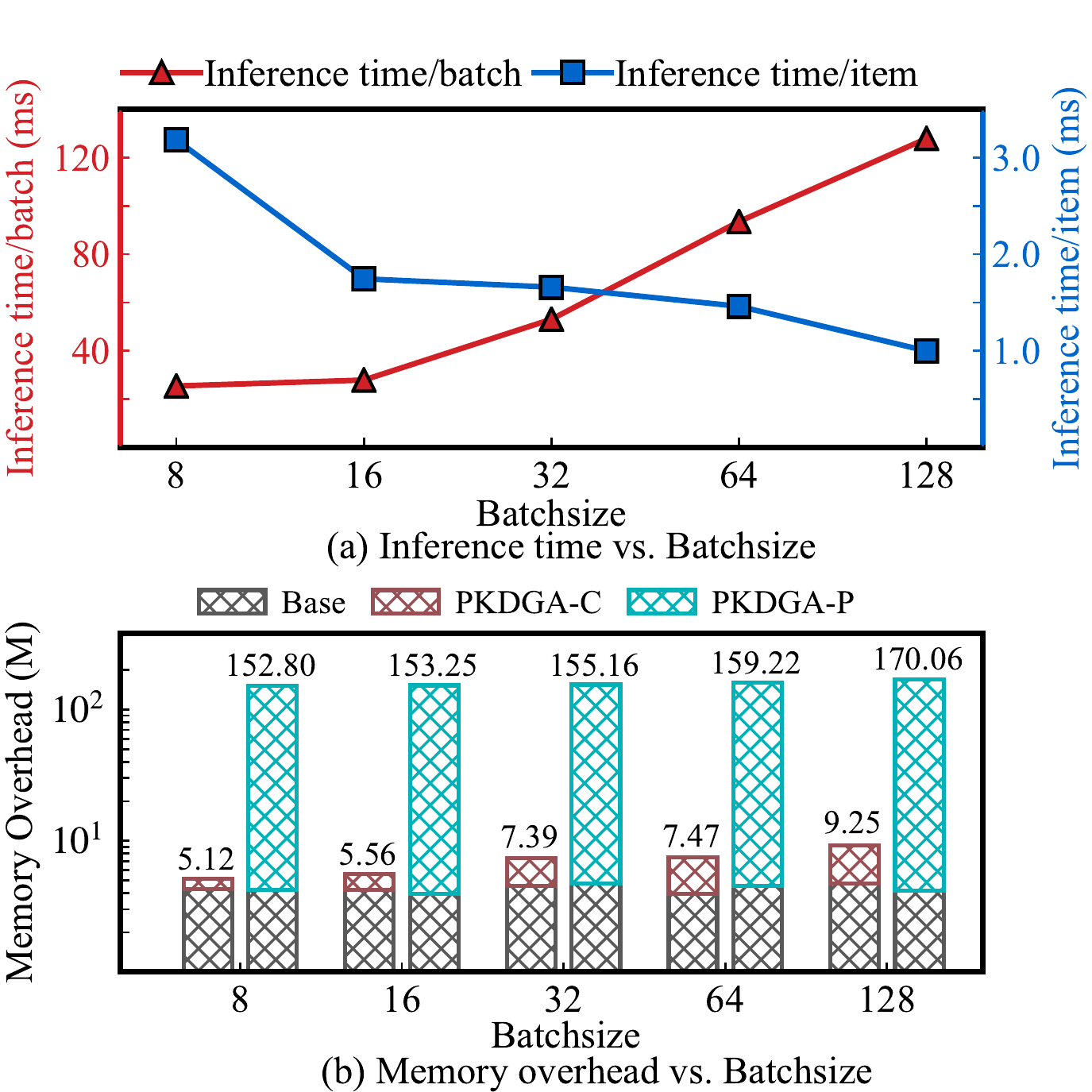}
	\vspace{-6pt}
	\caption{Overhead evaluation}
    \label{fig:overhead}
	\vspace{-16pt}
\end{figure}
We evaluate the \emph{PKDGA}'s overheads from two perspectives: (1) the training and inference time, and (2) the resource consumption when deploying \emph{PKDGA} in real-world malware. 

\textit{\textbf{Training time}}: We train \emph{PKDGA} on a machine equipped with an Intel(R) Xeon(R) Gold 6230 CPU with 128GB memory and an NVIDIA 2080Ti GPU. \emph{PKDGA} converges after 60 epochs, taking about 600 minutes. Since the training process occurs offline, this timespan is acceptable for adversaries.

\textit{\textbf{Inference time}}: The inference time is highly impacted by the batch size. We implement \emph{PKDGA} in \textit{Mirai} and evaluate the inference time using various batch sizes, and the results are shown in 
Fig.~\ref{fig:overhead}(a). The \textit{LSTM} network used by \emph{PKDGA} can generate a domain name in every 1-3 milliseconds. As we increase the batch size from 8 to 128, although the overall inference time increases from 25 milliseconds to 128 milliseconds, the average time consumed by generating a single domain instead decreases from 3.2 to 1 millisecond.



\textit{\textbf{Memory consumption}}:
We present the memory consumed by \emph{PKDGA} in Fig.~\ref{fig:overhead} (b), where the \emph{base} memory is occupied when solely running \textit{Mirai} without any DGA, \emph{PKDGA-P} and \emph{PKDGA-C} represent the two different implementations of \emph{PKDGA}: the \emph{C++ version} and the \emph{Python version}. 
As the \emph{Python version} needs to load a large number of dependency libraries before execution, its memory consumption is greater than 152.8MB. Compared with \emph{PKDGA-P}, \emph{PKDGA-C} consumes much less memory due to the simplified function and the elimination of the complex dependency libraries. The evaluations show that \emph{PKDGA-C} needs $30\times$ less memory. 

\vspace{-8pt}
\section{Discussion}\label{sec:dis}

Although \textit{PKDGA} enables adversaries to compromise AGD detection systems and helps botnets to avoid domain-related take-down risks, 
our ultimate goal in designing \textit{PKDGA} is to 
facilitate more advanced AGD detection strategies.
In other words, security vendors can avoid potential damage in advance by developing specific detection strategies that resist \textit{PKDGA} before adversaries realize reinforcement learning is an effective method of fooling DNS registration system.
However, defending against \textit{PKDGA} is challenging due to its adversarial characteristic, that is, \textit{PKDGA} can compromise arbitrary \textit{detectors} of specific status once feedback information is released. 
To this end, we propose a game-based strategy for detecting \textit{PKDGA}. The related motivation is described as follows. 

Training \textit{PKDGA} to compromise targets can be described as the process of exploring target's vulnerabilities, i.e., discovering the related adversarial AGDs. 
Hence, an intuitive way to defend against \textit{PKDGA} is to continually fix target's vulnerabilities until it is invulnerable. Here continually fixing vulnerabilities can be interpreted as incrementally learning to identify the novel adversarial AGDs. Inspired by this idea, we present our game-based AGD detection strategy as follows.
(1) Train \textit{PKDGA} to compromise target \textit{detector} $D_{target}$, 
{\setlength\abovedisplayskip{1pt}
\setlength\belowdisplayskip{2pt}
\begin{equation}
AGDs=PKDGA(D_{target}),
\label{eq:game}
\end{equation}
where $AGDs$ refer to the novel adversarial domains generated during training process. 
(2) Fix the vulnerabilities of $D_{target}$ by training it to incrementally identify the novel $AGDs$,}
{\setlength\abovedisplayskip{1pt}
\setlength\belowdisplayskip{2pt}
\begin{equation}
D^*_{target}=D_{target}(AGDs),
\end{equation}
where $D^*_{target}$ is the updated model after incremental learning.}
(3) Repeat steps (1)-(2) until \textit{PKDGA} cannot produce the novel adversarial AGDs against the target detector $D_{target}$.

\begin{figure}
	\centering
	\includegraphics[width=0.49\textwidth]{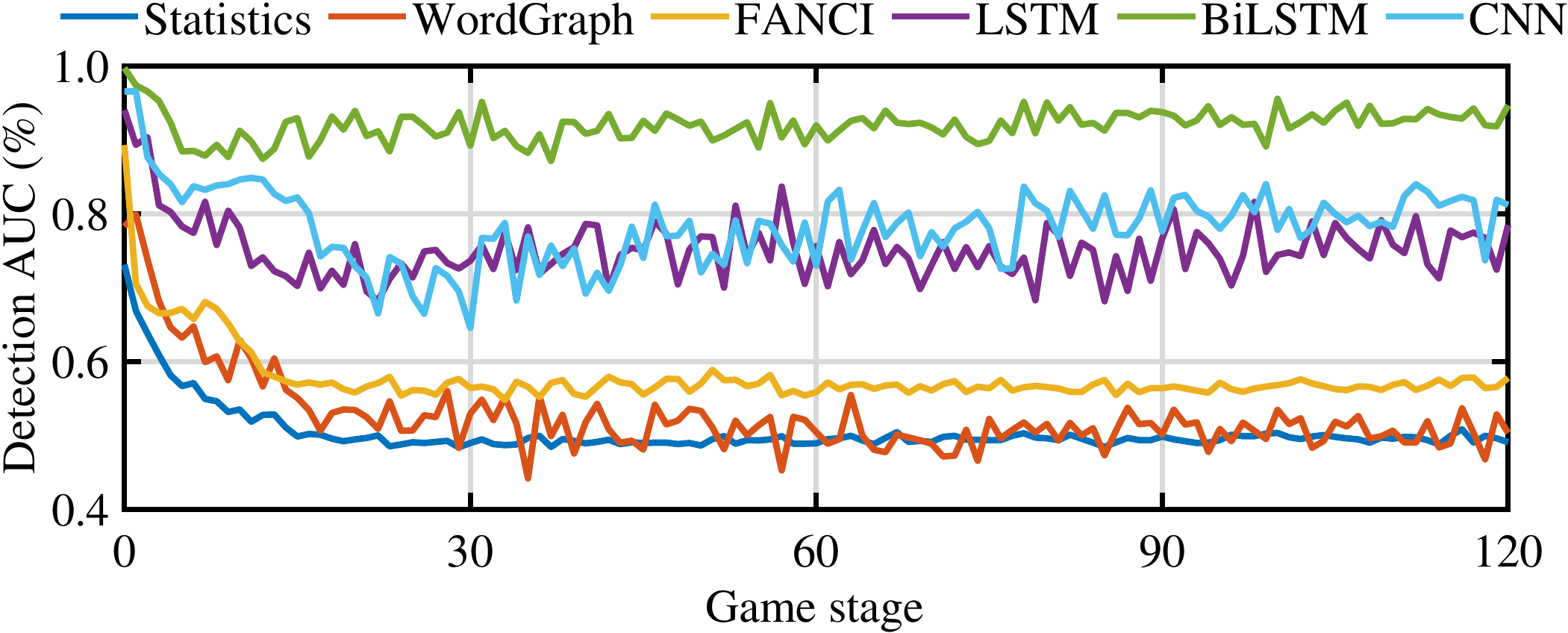}
	\vspace{-18pt}
	\caption{Game-based defending strategy}
	\vspace{-15pt}
    \label{fig:game}
\end{figure}

We modify the evaluated AGD \textit{detectors} into incremental versions and use them to implement this game-based detection strategy. The detection results are shown in Fig.~\ref{fig:game}, where a \textit{game stage} includes a step of \textit{PKDGA}  training and a step of \textit{detector} training. 
We can see that \textit{NN}-based \textit{detectors} (\textit{LSTM}, \textit{BiLSTM}, \textit{CNN}) achieve higher AUCs than the others.
This is because \textit{NN} models hold a more effective incremental update mechanism, tuning all of the parameters to identify novel samples. While the other \textit{detectors} can only update part or even none of their models, causing them to be ineffective when detecting \textit{PKDGA}. For instance, \textit{FANCI} can only update the leaf nodes or sub-tree to incrementally identify novel samples.

Although the game-based strategy achieves an $\ge$80\% detection AUC using incremental \textit{BiLSTM}, deploying such a defensive strategy in practice is still challenging 
since collecting the \textit{PKDGA} being used by an adversary will cost a large amount of time.
Security vendors must acquire adversaries' \textit{PKDGA} when generating novel AGDs (Eq.~\ref{eq:game}).
As far as we know, a feasible method of obtaining the in-use DGAs is to decompile the bot executable; however, this task it is very time-consuming and laborious. Decompiling a bot executable costs approximately tens of minutes, while the time of capturing the target bot executable is unpredictable or even infinite when the target executable is inaccessible. Moreover, security vendors must additionally re-capture the target bot executable once adversary updates \textit{PKDGA}, which renders this game-based defensive strategy expensive over the long term. Our future work will explore the game between incremental \textit{NN} \textit{detectors} and a local \textit{PKDGA} to reduce the time consumption caused by acquiring the target bot executable.

\vspace{-6pt}
\section{Related Work}\label{sec:rel}
\subsection{Domain Generation Algorithms}
We categorize the prior DGAs into two groups: zero-knowledge and full-knowledge.
Zero-knowledge DGAs~\cite{Szribi, CryptoLocker,Gameover-P2P, CharBot, Geodo, Gootkit} are widely used by real-life botnets~\cite{miris,Mirai,Kraken,Pykspa,Torpig,Conficker} since they do not make assumptions on \textit{detector} knowledge and produce AGDs using \textit{detector}-free algorithms. Although practical and lightweight, zero-knowledge DGAs are compromised by the latest \textit{detectors}~\cite{Distance-based, FANCI,LSTM}. Thus researchers propose full-knowledge DGAs~\cite{Khaos,MaskDGA,DeepDGA}, which assume the given information of the target \textit{detectors} and train learning-based generators to produce adversarial AGDs. For instance, \textit{Khaos}~\cite{Khaos} trains the \textit{Wasserstein generative adversarial network} (WGAN)~\cite{WGAN} to synthesize adversarial domains against a target \textit{detector}. However, \textit{Khaos} is challenged in terms of practicality since the security-sensitive information of targets is unavailable.

\vspace{-10pt}
\subsection{AGD detection methods}
The prior AGD detection methods can be roughly categorized into \textit{feature-based} and \textit{learning-based} types.

\textit{Feature-based methods} identify AGDs by extracting distinctive features on domain strings, where statistical and graph features are included.
\textit{Statistical features} (e.g., \textit{n-gram frequencies}~\cite{N-gram-distribution}) are designed to detect character-based AGDs, which have different distributions than the legitimate domains~\cite{Distance-based}. For example, \textit{Yadav et al.}~\cite{Distance-based} exploit distribution similarity (measured by the \textit{KL-divergence}~\cite{ref-20,ref-21,ref-10}, the \textit{Jaccard index}~\cite{ref-22} and the \textit{Edit distance}~\cite{ref-23}) to determine whether a query sample belongs to AGD groups or not. Although effective in detecting character-based DGAs, statistical features are challenged by the AGDs merging existing words. To this end, the \textit{graph feature} (\textit{WordGraph}~\cite{Graph_detection}) is proposed. In particular, \textit{WordGraph} constructs a graph describing relationships between all of the observed domains. The domains of degree larger than threshold are identified as AGDs. The rationale for this strategy is that word-based AGDs share the basic words sampled from a fixed dictionary~\cite{Gozi,Suppobox}, and the connections between those AGDs are closer (i.e., the corresponding degree in the relationship graph is larger) than the benign domains.

\textit{Learning-based methods}~\cite{ref-15,FANCI} integrate multiple features into learning models to improve the detection accuracy of single feature-based detection schemes. For instance, FANCI~\cite{FANCI} extracts 21 features of three types on domain strings and trains a random forest classifier to detect AGDs. Although effective in detecting AGDs of various types, its detection performance heavily relies on expert knowledge, which is required to manually extract the distinctive features. To this end, deep learning methods (e.g., \textit{CNN}~\cite{CNN}, \textit{LSTM}~\cite{LSTM}, \textit{NYU}~\cite{NYU}, \textit{MIT}~\cite{MIT} and \textit{LSTM*}~\cite{LSTMMI}) are proposed since they can use the raw data without feature engineering.

\vspace{-8pt}
\section{Conclusion}\label{sec:con}\vspace{-3pt}
DGAs enhance botnets by producing and registering dynamic domains for their C\&C servers. 
The existing AGD \textit{detectors} show considerably high accuracy when identifying the \emph{zero-knowledge DGAs}. \emph{Full-knowledge DGAs} that assume the full knowledge of target \textit{detectors} have limited practical ability in the real world. 
In this work, we propose \emph{PKDGA}, which uses only publicly available feedback to improve the anti-detection ability.
The experimental results show that it can achieve high anti-detection ability as well as high practicality. 
By applying \emph{PKDGA} to a botnet prototype system, we also
demonstrate that the proposed approach is time-efficient and lightweight in terms of memory and CPU overhead. \emph{PKDGA} can provide valuable information to existing machine learning-based security solutions and is able to contribute to a higher level of potential threat detection in a variety of environments.

\vspace{-10pt}
\bibliographystyle{IEEEtran}
\bibliography{usenix_security}

\begin{thebibliography}{10}
\providecommand{\url}[1]{#1}
\csname url@samestyle\endcsname
\providecommand{\newblock}{\relax}
\providecommand{\bibinfo}[2]{#2}
\providecommand{\BIBentrySTDinterwordspacing}{\spaceskip=0pt\relax}
\providecommand{\BIBentryALTinterwordstretchfactor}{4}
\providecommand{\BIBentryALTinterwordspacing}{\spaceskip=\fontdimen2\font plus
\BIBentryALTinterwordstretchfactor\fontdimen3\font minus
  \fontdimen4\font\relax}
\providecommand{\BIBforeignlanguage}[2]{{%
\expandafter\ifx\csname l@#1\endcsname\relax
\typeout{** WARNING: IEEEtran.bst: No hyphenation pattern has been}%
\typeout{** loaded for the language `#1'. Using the pattern for}%
\typeout{** the default language instead.}%
\else
\language=\csname l@#1\endcsname
\fi
#2}}
\providecommand{\BIBdecl}{\relax}
\BIBdecl

\bibitem{AGD_refer}
S.~Yadav, A.~K.~K. Reddy, A.~N. Reddy, and S.~Ranjan, ``Detecting
  algorithmically generated malicious domain names,'' in \emph{Proc. ACM
  SIGCOMM Conf. on Internet Meas.}, 2010, p. 48–61.

\bibitem{Khaos}
X.~Yun, J.~Huang, Y.~Wang, T.~Zang, Y.~Zhou, and Y.~Zhang, ``Khaos: An
  adversarial neural network dga with high anti-detection ability,'' \emph{IEEE
  Trans. Inf. Forensics Secur.}, vol.~15, no.~1, pp. 2225--2240, 2020.

\bibitem{Kraken}
M.~Zago, M.~G. P{\'e}rez, and G.~M. P{\'e}rez, ``Umudga: A dataset for
  profiling dga-based botnet,'' \emph{Comput. Secur.}, vol.~92, p. 101719,
  2020.

\bibitem{DGArchive}
D.~Plohmann, K.~Yakdan, M.~Klatt, J.~Bader, and E.~Gerhards-Padilla, ``A
  comprehensive measurement study of domain generating malware,'' in
  \emph{Proc. USENIX Secur. Symp.}, 2016, pp. 263--278.

\bibitem{Gozi}
\BIBentryALTinterwordspacing
E.~Team, ``Tracking rovnix,'' Blog post, 2014. [Online]. Available:
  \url{https://www.bitdefender.com/blog/labs/tracking-rovnix-2/}
\BIBentrySTDinterwordspacing

\bibitem{Pykspa}
\BIBentryALTinterwordspacing
J.~BADER, ``Domain generation algorithm analyses,'' Blog posts, 2015. [Online].
  Available: \url{https://bin.re/tag/dga/}
\BIBentrySTDinterwordspacing

\bibitem{Suppobox}
J.~Geffner, ``End-to-end analysis of a domain generating algorithm malware
  family,'' in \emph{Proc. Blackhat Conf.}, 2013.

\bibitem{take-down-1}
F.~Casino, N.~Lykousas, I.~Homoliak, C.~Patsakis, and J.~Hernandez-Castro,
  ``Intercepting hail hydra: Real-time detection of algorithmically generated
  domains,'' \emph{J. Netw. Comput. Appl.}, vol. 190, p. 103135, 2021.

\bibitem{bot_take}
V.~Le~Pochat, S.~Maroofi, T.~Van~Goethem, D.~Preuveneers, A.~Duda, W.~Joosen,
  Korczy{\'n}ski, and Maciej, ``A practical approach for taking down avalanche
  botnets under real-world constraints,'' in \emph{Proc. Netw. Distrib. Syst.
  Secur. Symp.}, 2020.

\bibitem{FANCI}
S.~Sch{\"u}ppen, D.~Teubert, P.~Herrmann, and U.~Meyer, ``{FANCI} :
  Feature-based automated nxdomain classification and intelligence,'' in
  \emph{Proc. USENIX Secur. Symp.}, 2018, pp. 1165--1181.

\bibitem{TDSC-takedown}
Y.~Nadji, R.~Perdisci, and M.~Antonakakis, ``Still beheading hydras: Botnet
  takedowns then and now,'' \emph{IEEE Trans. on Dependable and Secur.
  Comput.}, vol.~14, no.~5, pp. 535--549, 2017.

\bibitem{Ransomware}
S.~Mohurle and M.~Patil, ``A brief study of wannacry threat: Ransomware attack
  2017,'' \emph{Int. J. of Adv. Res. in Comput. Sci.}, vol.~8, no.~5, pp.
  1938--1940, 2017.

\bibitem{spam-campaigns}
H.~Gao, J.~Hu, C.~Wilson, Z.~Li, Y.~Chen, and B.~Y. Zhao, ``Detecting and
  characterizing social spam campaigns,'' in \emph{Internet Meas. Conf.}, 2010,
  pp. 35--47.

\bibitem{DDoS}
F.~Lau, S.~H. Rubin, M.~H. Smith, and L.~Trajkovic, ``Distributed denial of
  service attacks,'' in \emph{Int. Conf. on Syst., Man, and Cybern.},
  vol.~3.\hskip 1em plus 0.5em minus 0.4em\relax IEEE, 2000, pp. 2275--2280.

\bibitem{Graph_detection}
M.~Pereira, S.~Coleman, B.~Yu, M.~DeCock, and A.~Nascimento, ``Dictionary
  extraction and detection of algorithmically generated domain names in passive
  dns traffic,'' in \emph{Proc. Int. Symp. Res. in Attacks, Intrusions, and
  Defenses}, 2018, pp. 295--314.

\bibitem{Distance-based}
S.~Yadav, A.~K.~K. Reddy, A.~L.~N. Reddy, and S.~Ranjan, ``Detecting
  algorithmically generated domain-flux attacks with dns traffic analysis,''
  \emph{IEEE/ACM Trans. Netw.}, vol.~20, no.~5, pp. 1663--1677, 2012.

\bibitem{adversarial_example_1}
N.~Carlini and D.~Wagner, ``Towards evaluating the robustness of neural
  networks,'' in \emph{Proc. IEEE Symp. on Secur. and Privacy}, 2017, pp.
  39--57.

\bibitem{adversarial_example_2}
S.-M. Moosavi-Dezfooli, A.~Fawzi, O.~Fawzi, and P.~Frossard, ``Universal
  adversarial perturbations,'' in \emph{Proc. IEEE Conf. Comput. Vis. and
  Pattern Recognit.}, 2017, pp. 1765--1773.

\bibitem{AWA}
A.~M. Sadeghzadeh, B.~Tajali, and R.~Jalili, ``Awa: Adversarial website
  adaptation,'' \emph{IEEE Trans. Inf. Forensics Secur.}, vol.~16, no.~1, pp.
  3109--3122, 2021.

\bibitem{MaskDGA}
L.~Sidi, A.~Nadler, and A.~Shabtai, ``Maskdga: An evasion attack against dga
  classifiers and adversarial defenses,'' \emph{IEEE Access}, vol.~8, pp.
  161\,580--161\,592, 2020.

\bibitem{Mirai}
M.~Antonakakis, T.~April, M.~Bailey, M.~Bernhard, E.~Bursztein, J.~Cochran,
  Z.~Durumeric, J.~A. Halderman, L.~Invernizzi, M.~Kallitsis \emph{et~al.},
  ``Understanding the mirai botnet,'' in \emph{Proc. USENIX Secur. Symp.},
  2017, pp. 1093--1110.

\bibitem{Kraken1}
R.~Behrends, L.~K. Dillon, S.~D. Fleming, and R.~E.~K. Stirewalt, ``On the
  kraken and bobax botnets,'' \emph{Tech. rep.}, 2008.

\bibitem{CNN}
J.~Saxe and K.~Berlin, ``expose: A character-level convolutional neural network
  with embeddings for detecting malicious urls, file paths and registry keys,''
  \emph{arXiv}, 2017.

\bibitem{LSTM}
J.~Woodbridge, H.~S. Anderson, A.~Ahuja, and D.~Grant, ``Predicting domain
  generation algorithms with long short-term memory networks,'' \emph{arXiv},
  2016.

\bibitem{AUC}
J.~A. Hanley and B.~J. McNeil, ``The meaning and use of the area under a
  receiver operating characteristic (roc) curve.'' \emph{Radiology}, vol. 143,
  no.~1, pp. 29--36, 1982.

\bibitem{NYU}
X.~Zhang, J.~Zhao, and Y.~LeCun, ``Character-level convolutional networks for
  text classification,'' in \emph{Adv. in Neural Inf. Process. Syst.}, vol.~28,
  2015, pp. 649--657.

\bibitem{MIT}
S.~Vosoughi, P.~Vijayaraghavan, and D.~Roy, ``Tweet2vec: Learning tweet
  embeddings using character-level cnn-lstm encoder-decoder,'' in \emph{Proc.
  Int. ACM SIGIR Conf. on Res. and Develop. in Inf. Retrieval}, 2016, p.
  1041–1044.

\bibitem{CMU}
B.~Dhingra, Z.~Zhou, D.~Fitzpatrick, M.~Muehl, and W.~Cohen, ``Tweet2vec:
  Character-based distributed representations for social media,'' in
  \emph{Proc. Annu. Meeting of the Assoc. for Comput. Linguistics}, 2016, pp.
  269--274.

\bibitem{LSTMMI}
D.~Tran, H.~Mac, V.~Tong, H.~A. Tran, and L.~G. Nguyen, ``A lstm based
  framework for handling multiclass imbalance in dga botnet detection,''
  \emph{Neurocomputing}, vol. 275, pp. 2401--2413, 2018.

\bibitem{Bamital}
P.~Krysiuk and V.~Thakur, ``Trojan. bamital,'' \emph{Tech. rep.}, 2013.

\bibitem{Bedep}
\BIBentryALTinterwordspacing
D.~Schwarz, ``Bedep’s dga: Trading foreign exchange for malware domains,''
  Blog post, 2016. [Online]. Available:
  \url{https://asert.arbornetworks.com/bedeps-dga-trading-foreign-exchange-for-malware-domains/}
\BIBentrySTDinterwordspacing

\bibitem{Conficker}
F.~Leder and T.~Werner, ``Know your enemy: Containing conficker, to tame a
  malware,'' \emph{Tech. rep.}, 2009.

\bibitem{Gootkit}
\BIBentryALTinterwordspacing
U.~Parasites, ``Runforestrun and pseudo random domains,'' Blog post, 2012.
  [Online]. Available: \url{http://blog.unmaskparasites.com/2012/06/22/
  runforestrun-and-pseudo-random-domains/}
\BIBentrySTDinterwordspacing

\bibitem{Hesperbot}
A.~Cherepanov and R.~Lipovsky, ``Hesperbot-a new, advanced banking trojan in
  the wild,'' 2013.

\bibitem{Szribi}
\BIBentryALTinterwordspacing
J.~Wolf, ``Technical details of srizbi’s domain generation algorithm,'' Blog
  post, 2008. [Online]. Available:
  \url{https://forum.security-x.fr/news/(fireeye)technical-details-of-srizbi039s-domain-generation-algorithm/}
\BIBentrySTDinterwordspacing

\bibitem{CryptoLocker}
K.~Liao, Z.~Zhao, A.~Doup{\'e}, and G.-J. Ahn, ``Behind closed doors:
  measurement and analysis of cryptolocker ransoms in bitcoin,'' in \emph{APWG
  symp. on electron. crime res.}, 2016, pp. 1--13.

\bibitem{Torpig}
B.~Stone-Gross, M.~Cova, L.~Cavallaro, B.~Gilbert, M.~Szydlowski, R.~Kemmerer,
  C.~Kruegel, and G.~Vigna, ``Your botnet is my botnet: analysis of a botnet
  takeover,'' in \emph{Proc. ACM Conf. on Comput. and commun. secur.}, 2009,
  pp. 635--647.

\bibitem{Gameover-P2P}
D.~Andriesse, C.~Rossow, B.~Stone-Gross, and D.~Plohmann, ``Highly resilient
  peer-to-peer botnets are here: An analysis of gameover zeus,'' in \emph{Int.
  Conf. on Malicious and Unwanted Softw.}, 2013, pp. 116--123.

\bibitem{CharBot}
J.~Peck, C.~Nie, R.~Sivaguru, C.~Grumer, F.~Olumofin, B.~Yu, and A.~Nascimento,
  ``Charbot: A simple and effective method for evading dga classifiers,''
  \emph{IEEE Access}, vol.~7, no.~1, pp. 91\,759--91\,771, 2019.

\bibitem{Geodo}
\BIBentryALTinterwordspacing
M.~P. CENTER, ``Trojan:win32/emotet.c,'' Blog post, 2014. [Online]. Available:
  \url{https://www.microsoft.com/security/portal/threat/ encyclopedia/entry.
  aspx?Name=Trojan:Win32/Emotet.C.}
\BIBentrySTDinterwordspacing

\bibitem{Nymaim}
\BIBentryALTinterwordspacing
T.~Barabosch and E.~Gerhards-Padilla, ``Behavior-driven development in malware
  analysis,'' Blog post, 2015. [Online]. Available:
  \url{https://itsec.cs.uni-bonn.de/spring2015/downloads/barabosch.pdf}
\BIBentrySTDinterwordspacing

\bibitem{DeepDGA}
H.~S. Anderson, J.~Woodbridge, and B.~Filar, ``Deepdga: Adversarially-tuned
  domain generation and detection,'' in \emph{Proc. ACM Workshop on Artif.
  Intell. and Secur.}, 2016, pp. 13--21.

\bibitem{COMST_RL}
X.~Wang, Y.~Han, V.~C.~M. Leung, D.~Niyato, X.~Yan, and X.~Chen, ``Convergence
  of edge computing and deep learning: A comprehensive survey,'' \emph{IEEE
  Commun. Surv. Tut.}, vol.~22, no.~2, pp. 869--904, 2020.

\bibitem{SeqGAN}
L.~Yu, W.~Zhang, J.~Wang, and Y.~Yong, ``Seqgan: Sequence generative
  adversarial nets with policy gradient,'' in \emph{Proc. AAAI Conf. Artif.
  Intell.}, 2017.

\bibitem{Policy-Gradient}
R.~S. Sutton, D.~Mcallester, S.~Singh, and Y.~Mansour, ``Policy gradient
  methods for reinforcement learning with function approximation,'' in
  \emph{Proc. Adv. Neural Inf. Process. Syst.}, 1999.

\bibitem{deeplog}
M.~Du, F.~Li, G.~Zheng, and V.~Srikumar, ``Deeplog: Anomaly detection and
  diagnosis from system logs through deep learning,'' in \emph{Proc. ACM SIGSAC
  Conf. Comput. and Commun. Secur.}, 2017, pp. 1285--1298.

\bibitem{inter_speech}
T.~Mikolov, M.~Karafi{\'a}t, L.~Burget, J.~Cernock{\`y}, and S.~Khudanpur,
  ``Recurrent neural network based language model.'' in \emph{Interspeech},
  2010, pp. 1045--1048.

\bibitem{LSTM_pro}
R.~Jozefowicz, W.~Zaremba, and I.~Sutskever, ``An empirical exploration of
  recurrent network architectures,'' in \emph{Int. conf. on mach. learn.}\hskip
  1em plus 0.5em minus 0.4em\relax PMLR, 2015, pp. 2342--2350.

\bibitem{LSTM_intro}
S.~Hochreiter and J.~Schmidhuber, ``Long short-term memory,'' \emph{Neural
  Comput.}, vol.~9, no.~8, pp. 1735--1780, 1997.

\bibitem{PKDGA}
``https://github.com/sxy1017/pkdga.''

\bibitem{DNS}
``https://thekelleys.org.uk/dnsmasq/.''

\bibitem{Alexa}
``https://www.alexa.com/topsites,'' 2021.

\bibitem{WGAN}
J.~Adler and S.~Lunz, ``Banach wasserstein gan,'' in \emph{Proc. Adv. Neural
  Inf. Process. Syst.}, 2018.

\bibitem{FlowPrint}
T.~van Ede, R.~Bortolameotti, A.~Continella, J.~Ren, D.~J. Dubois,
  M.~Lindorfer, D.~Choffnes, M.~van Steen, and A.~Peter, ``Flowprint:
  Semi-supervised mobile-app fingerprinting on encrypted network traffic,'' in
  \emph{Proc. Netw. Distrib. Syst. Secur. Symp.}, 2020, pp. 1--18.

\bibitem{TNSM}
L.~Nie, L.~Zhao, and K.~Li, ``Robust anomaly detection using reconstructive
  adversarial network,'' \emph{IEEE Trans. Netw. Serv. Manag.}, vol.~18, no.~2,
  pp. 1899--1912, 2021.

\bibitem{adversarial_example}
N.~Carlini and D.~Wagner, ``Towards evaluating the robustness of neural
  networks,'' in \emph{Proc. IEEE Symp. on Secur. and Privacy}, 2017, pp.
  39--57.

\bibitem{miris}
\BIBentryALTinterwordspacing
C.~Osborne, ``Meris botnet assaults krebs on security: The botnet appears to be
  made up of compromised routers,'' Blog post, 2021. [Online]. Available:
  \url{https://www.zdnet.com/article/ meris-botnet-assaults-krebsonsecurity}
\BIBentrySTDinterwordspacing

\bibitem{N-gram-distribution}
M.~Antonakakis, R.~Perdisci, Y.~Nadji, N.~Vasiloglou, S.~Abu-Nimeh, W.~Lee, and
  D.~Dagon, ``From throw-away traffic to bots: Detecting the rise of dga-based
  malware,'' in \emph{Proc. USENIX Secur. Symp.}, 2012, p.~24.

\bibitem{ref-20}
S.~Kullback and R.~A. Leibler, ``On information and sufficiency,'' \emph{Ann.
  Math. Statist.}, vol.~22, no.~1, pp. 79--86, 1951.

\bibitem{ref-21}
S.~Kullback, \emph{Information theory and statistics}.\hskip 1em plus 0.5em
  minus 0.4em\relax Courier Corporation, 1997.

\bibitem{ref-10}
Y.~Fu, L.~Yu, O.~Hambolu, I.~Ozcelik, B.~Husain, J.~Sun, K.~Sapra, D.~Du, C.~T.
  Beasley, and R.~R. Brooks, ``Stealthy domain generation algorithms,''
  \emph{IEEE Trans. Inf. Forensics Secur.}, vol.~12, no.~6, pp. 1430--1443,
  2017.

\bibitem{ref-22}
V.~I. Levenshtein \emph{et~al.}, ``Binary codes capable of correcting
  deletions, insertions, and reversals,'' in \emph{Sov. phys. doklady},
  vol.~10, no.~8.\hskip 1em plus 0.5em minus 0.4em\relax Soviet Union, 1966,
  pp. 707--710.

\bibitem{ref-23}
H.~Small, ``Co-citation in the scientific literature: A new measure of the
  relationship between two documents,'' \emph{J. of the Amer. Soc. for inf.
  Sci.}, vol.~24, no.~4, pp. 265--269, 1973.

\bibitem{ref-15}
S.~Schiavoni, F.~Maggi, L.~Cavallaro, and S.~Zanero, ``Phoenix: Dga-based
  botnet tracking and intelligence,'' in \emph{Int. Conf. on Detection of
  Intrusions and Malware, and Vulnerability Assessment}.\hskip 1em plus 0.5em
  minus 0.4em\relax Springer, 2014, pp. 192--211.

\end{thebibliography}
\vspace{-35pt}
\begin{IEEEbiographynophoto}{Lihai Nie}
received the bachelor's degree and master's degree from the Dalian Maritime University, China, in 2015 and 2018. He is currently working toward the PhD degree at College of Intelligence and Computing, Tianjin University, China. His research interests include network security and machine learning.
\end{IEEEbiographynophoto}
\vspace{-35pt}
\begin{IEEEbiographynophoto}{Xiaoyang Shan}
received the bachelor's degree from the Xidian University, China, in 2020. She is currently working toward the master's degree at College of Intelligence and Computing, Tianjin University, China. Her research interests include cybersecurity and machine learning.
\end{IEEEbiographynophoto}
\vspace{-35pt}
\begin{IEEEbiographynophoto}{Laiping Zhao}
received the BS and MS degrees from Dalian University of Technology, China, in 2007 and 2009, and the PhD degree from the Department of Informatics, Kyushu University, Japan, in 2012. He is currently an associate professor with the School of Computer Software, Tianjin University, China. His research interests include cloud computing and cybersecurity.
\end{IEEEbiographynophoto}
\vspace{-35pt}
\begin{IEEEbiographynophoto}{Keqiu Li}
received the bachelor's and master's degrees from the Department of Applied Mathematics, Dalian University of Technology, in 1994 and 1997, respectively, and the PhD degree from the Graduate School of Information Science, Japan Advanced Institute of Science and Technology, in 2005. He is currently a professor in the College of Intelligence \& Computing, Tianjin University, China. He has published more than 100 technical papers, such as the IEEE Transactions on Parallel and Distributed Systems, the ACM Transactions on Internet Technology, and the ACM Transactions on Multimedia Computing, Communications, and Applications. He is an associate editor of the
IEEE Transactions on Parallel and Distributed Systems and the IEEE Transactions on Computers. His research interests include data center networks, cloud computing, and cybersecurity. He is a fellow of the IEEE.
\end{IEEEbiographynophoto}
\end{document}